\documentclass[preprint,12pt]{elsarticle}
\biboptions{numbers,sort&compress}

\newcounter{bla}

\journal{Computer Physics Communications}

\usepackage{graphicx}
\usepackage{dcolumn}
\usepackage{bm}
\usepackage{amsmath}
\usepackage{amssymb}
\usepackage{amsfonts}
\usepackage{subfigure}
\usepackage{esvect}
\usepackage[dvipsnames]{xcolor}
\usepackage{fix-cm} %allows large fonts
\usepackage{soul}
\usepackage{mdframed}
\usepackage{lipsum}
\usepackage{multicol}
\usepackage{float}
\usepackage{longtable}
\usepackage{setspace}
\usepackage{tabularx}
\usepackage{ltablex}
\usepackage{xspace}
\usepackage{newfloat}
\DeclareFloatingEnvironment[fileext=frm,placement={!ht},name=Box]{boxfloat}
\usepackage{capt-of}
\usepackage{hyperref}
\usepackage{url}
\usepackage{eso-pic}

\usepackage{tikz}
\usetikzlibrary{decorations.pathreplacing}

\usepackage{listings}
\usepackage{xcolor}
\usepackage{listings}
\lstdefinestyle{cppstyle}{
  language=C++,
  basicstyle=\ttfamily,
  keywordstyle=\color{black},
  numbers=none,
  frame=none,
  columns=fullflexible,
  keepspaces=true,
  showstringspaces=false,
  aboveskip=0.5em,
  belowskip=0.5em,
  emph={double},
  emphstyle=\color{magenta!80!black},
  emph={[2]auto},
  emphstyle={[2]\color{green!50!black}
  }
}
\lstdefinestyle{terminal}{
  basicstyle=\ttfamily\color{gray!80!black},
  numbers=none,
  frame=tb,
  rulecolor=\color{gray!70!black},
  framerule=0.4pt,
  columns=fullflexible,
  keepspaces=true,
  showstringspaces=false,
  aboveskip=0.7em,
  belowskip=0.7em,
  xleftmargin=0pt,
  framexleftmargin=0pt
}
\usepackage{minted}
\newminted{shell-session}{
    frame=lines, breaklines=true}
\definecolor{mintbackground}{RGB}{246, 246, 246}

\usepackage[symbol]{footmisc}

\newcommand{\CL}[1]{$\mathcal{CL}\mathtt{\,\,v#1}$}

\newcommand{\CLns}{$\mathcal{CL}$}

\newcommand{\cosmolattice}{$\tt {\mathcal C}osmo{\mathcal L}attice~$}

\newcommand{\ArtI}{{\tt The Art\,-\,I}}
\newcommand{\ArtII}{{\tt The Art\,-\,II}}
\newcommand{\ArtIII}{{\tt The Art\,-\,III}}
\newcommand{\ArtIV}{{\tt The Art\,-\,IV}}

\newcommand{\dx}{\ensuremath{\delta x}}

\newcommand{\be}{\begin{equation}}
    \newcommand{\ee}{\end{equation}}
\newcommand{\bea}{\begin{eqnarray}}
    \newcommand{\eea}{\end{eqnarray}}

\newcommand{\TempLat}{\texttt{TempLat}\xspace}
\newcommand{\ParaFaFT}{\texttt{ParaFaFT}\xspace}

\begin{document}

\AddToShipoutPictureFG*{%
  \put(\LenToUnit{20cm},\LenToUnit{27cm}){%
    \makebox(0,0)[rt]{\small MS-TP-26-22}%
    \makebox(0,-13)[rt]{\small DESY-26-104}%
  }%
}

\begin{frontmatter}

\title{-- \cosmolattice {\tt 2.0} -- \\ A modern code for lattice simulations of scalar \\and gauge field dynamics in an expanding universe}

%% use optional labels to link authors explicitly to addresses:
%% \author[label1,label2]{<author name>}
%% \address[label1]{<address>}
%% \address[label2]{<address>}

\author[a]{Jorge Baeza-Ballesteros}
\author[b]{Daniel~G.~Figueroa}
\author[c]{Adrien Florio}
\author[d]{Nicol\'as Loayza}
\author[c]{\\Franz R.~Sattler}
\author[e]{Francisco Torrent\'i}
\author[f,g,h]{Ander Urio}

%\cortext[author3] {\\\textit{E-mail address:} daniel.figueroa@ific.uv.es}
%\cortext[author2] {\\\textit{E-mail address:} adrien.florio@stonybrook.edu}
%\cortext[author3] {\\\textit{E-mail address:} f.torrenti@unibas.ch}
\address[a]{Deutsches Elektronen-Synchrotron DESY, Platanenallee 6, 15738 Zeuthen, Germany}
\address[b]{Instituto de F\'isica Corpuscular (IFIC), Universitat de Val\`encia (UV) and\\
\it Consejo Superior de Investigaciones Cient\'ificas (CSIC), 46980 Paterna, Valencia, Spain}
\address[c]{Fakult\"at f\"ur Physik, Universit\"at Bielefeld, D-33615 Bielefeld, Germany}
\address[d]{CEICO-FZU, Institute of Physics of the Czech Academy of Sciences, Na Slovance 1999/2, 182 00, Prague, Czechia}
\address[e]{Department of Mathematics, Universidad Carlos III de Madrid, Avenida de la Universidad 30, 28911 Legan\'es, Madrid, Spain}
\address[f]{Department of Physics, University of the Basque Country, UPV/EHU, 48080, Bilbao, Spain}
\address[g]{Department of Applied Mathematics, University of the Basque Country UPV/EHU, Plaza Ingeniero Torres Quevedo 1, 48013 Bilbao, Spain}
\address[h]{EHU Quantum Center, University of the Basque Country UPV/EHU, Leioa, 48940 Biscay, Spain}

%%%%%%%%%%%%%%%%%%%%%%%%%%%%%%%%%%%%%%%%%%%%%%%%%%%%%%%%%%%%%%%%%%%%%%%%%%%%%%%%

\begin{abstract}
This paper introduces \cosmolattice %(\CLns) 
{\tt v2.0}, a major upgrade that substantially broadens the physical scope and computational capabilities of the code. It introduces lattice implementations of scalar fields non-minimally coupled to gravity through $\phi^2R$, as well as axion-like fields coupled to Abelian gauge sectors as $\phi F_{\mu\nu}\widetilde F^{\mu\nu}$. It also provides new procedures for generating specialized initial conditions, including scaling networks of cosmic defects ({\it e.g.}~strings and domain walls), and fields with arbitrary power spectra. The release also incorporates low-storage Runge--Kutta integrators for non-symplectic systems (suitable {\it e.g.}~for non-minimal scalar kinetic terms as $\mathcal{G}_{ab}\partial_\mu\phi^a\partial^\mu\phi^b$), scalar-field simulations on reduced $(1+1)$- and $(2+1)$-dimensional lattices, new optimized gravitational-wave evolution, more flexible field and energy-density outputs, and GPU support that can accelerate simulations by a factor $\mathcal{O}(10)$ relative to CPU execution. Extensive documentation on the use of the code is provided on \href{http://www.cosmolattice.com}{\color{blue}http://www.cosmolattice.com}\,.
\end{abstract}

%%%%%%%%%%%%%%%%%%%%%%%%%%%%%%%%%%%%%%%%%%%%%%%%%%%%%%%%%%%%%%%%%%%%%%%%%%%%%%%%

%\begin{keyword}
%early Universe, real-time lattice simulations, non-linear field dynamics, scalar singlet interactions, scalar-gauge interactions, axion-gauge interactions, non-minimal interactions, gauge-invariant lattice techniques.
%\end{keyword}

\end{frontmatter}

\tableofcontents  %\newpage
%{\em Additional comments including restrictions and unusual features (approx. 50-250 words):}\\
%  %Provide any additional comments here.

%%%%%%%%%%%%%%%%%%%%%%%%%%%%%%%%%%%%%%%%%%%%%%%%%%%%%%%%%%%%%%%%%%%%%%%%%%%%%%%%
\section{Introduction}
\label{sec:Intro}

The application of numerical methods to the study of non-linear field dynamics in the early Universe has grown substantially, as reflected {\it e.g.}~by the number of specialized packages developed in recent years~\cite{Figueroa:2021yhd,Figueroa:2023xmq,Daverio:2015ryl,Lozanov:2019jff,Giblin:2019nuv,Andrade:2021rbd,Buschmann:2024bfj,Caravano:2025klk}, see also~\cite{Felder:2000hq,Felder:2007nz,Frolov:2008hy,Sainio:2009hm,Easther:2010qz,Huang:2011gf,Sainio:2012mw}. Building increasingly efficient, accurate and robust numerical tools is essential for placing predictions about the early Universe on a firm theoretical foundation. These developments have led to the emergence of a distinct field---{\bf Lattice Cosmology}---whose strength lies in its ability to resolve the detailed non-linear dynamics of fields, providing reliable predictions for the resulting observables of primordial phenomena. {\bf Lattice Cosmology Techniques} (LCT) nowadays represent a well-established route towards understanding the physics of the early Universe, and are expected to play an increasingly prominent role in shaping future observational strategies aimed at probing this largely unexplored epoch.

It was in this context that \cosmolattice \cite{Figueroa:2021yhd} was originally developed, as a dedicated software to solve the non-linear dynamics of field theories involving scalar and gauge fields embedded in an expanding background. Rather than being a conventional code, designed to solve a predetermined set of equations and compute a fixed collection of observables, \cosmolattice constitutes a flexible {\it platform} for implementing field-theory systems governed by partial differential equations that can be discretized on a lattice~\cite{Figueroa:2023xmq}. Written in C++, \cosmolattice follows a modular architecture that cleanly separates the  physics from the underlying technical implementation. It also introduces a dedicated symbolic language in which field variables and the operations acting on them are expressed in a form that closely resembles their continuum counterparts.

Released in 2021, \cosmolattice {\tt v1.0}~\cite{Figueroa:2021yhd} provided a public tool to simulate {\it canonical} scalar-singlet and $\mathrm{SU}(2)$$\times$$\mathrm{U}(1)$ scalar-gauge theories in a spatially-flat expanding Universe. In particular, %{\tt v1.0} of the code 
it incorporated explicit-time symplectic evolution algorithms for theories with canonically normalized kinetic terms, with accuracies ranging from $\mathcal{O}(\delta t^{2})$ to $\mathcal{O}(\delta t^{10})$. It also supported self-consistent expansion of the Universe sourced by all scalar and gauge fields, while preserving lattice gauge invariance, and consequently the Gauss constraints of both U(1) and SU(2) sectors to machine precision  throughout the evolution. In this paper, we present \cosmolattice {\tt v2.0}, which constitutes a substantial upgrade to the code, introducing a broad range of new capabilities. These include: {\bf a)} lattice implementations of non-canonical interactions, like scalars with a non-minimal coupling to gravity $\phi^2R$, or axion-like particles (ALP) interacting with gauge fields $\phi F_{\mu\nu}\tilde F^{\mu\nu}$; {\bf b)} methods to set up special field configurations, like cosmic defect networks close to the {\it scaling regime} ({\it e.g.}~cosmic strings and domain walls), or initial conditions from arbitrary power spectra; {\bf c)} new technical features, such as non-symplectic evolution algorithms (suitable {\it e.g.} for non-minimal scalar kinetic theories of the type $\mathcal{G}_{ab}(\lbrace\phi_c\rbrace)\partial_\mu\phi^a\partial^\mu\phi^b$), optimized implementations of gravitational wave (GW) dynamics on the lattice (sourced by scalar and gauge fields), scalar field dynamics in $1 + 1$ and $2+1$ dimensions, and others; and {\bf d)} a fully reworked backend and GPU support, gaining a factor $\mathcal{O}(10)$ in performance as compared to running on CPUs. 

In this paper, we explain the new capabilities of \cosmolattice {\tt v2.0}. In Sect.~\ref{sec:previousCL} we review the inherited capabilities from previous versions of the code, whereas in Sects.~\ref{sec:NewPhysics},~\ref{sec:NewFeatures}, and~\ref{sec:NewCapabilities}, we discuss the new physics, features, and performance capabilities, respectively, that \cosmolattice {\tt v2.0} incorporates. Finally, we conclude in Sect.~\ref{sec:Discussion} and provide an outlook on future development directions. An extensive documentation on the use of the code is provided on the \href{http://www.cosmolattice.com}{\color{blue} \cosmolattice website}. The %latest version of the 
code can be downloaded from there, or from its \href{https://github.com/cosmolattice/cosmolattice}{\color{blue} GitHub repository}.

{\bf Note --.} The release of \cosmolattice {\tt v1.0} was accompanied by a monographic review on LCT: {\it The Art of Simulating the early Universe, Part I. Integration techniques and canonical cases}~\cite{Figueroa:2020rrl}, which constituted the theoretical foundation of {\tt v1.0} of the code. We refer colloquially to that review as {\tt The Art\,-\,I}. The release of \cosmolattice {\tt v2.0} with the present publication, is also accompanied by another monographic review on LCT: {\it The Art of Simulating the early Universe, Part II. Non-canonical cases \& gravitational waves}~\cite{Baeza-Ballesteros:2025tme} or, for short, {\tt The Art\,-\,II}, where the new aspects incorporated in \cosmolattice {\tt v2.0} are discussed at a theoretical level. In this paper, we will often refer to {\tt The Art\,-\,I}~\cite{Figueroa:2020rrl} or {\tt The Art\,-\,II}~\cite{Baeza-Ballesteros:2025tme} for theoretical clarifications.

%%%%%%%%%%%%%%%%%%%%%%%%%%%%%%%%%%%%%%%%%%%%%%%%%%%%%%%%%%%%%%%%%%%%%%%%%%%%%%%%
\section{Inherited capabilities from \CL{1.X}}
\label{sec:previousCL}

We first review the field sectors that previous versions of \cosmolattice can simulate, as version {\tt 2.0} of the code inherits them. All versions {\tt 1.X} -- from now on \CL{1.X} \mbox{--,} can solve the dynamics of \textit{canonical} field theories embedded in an expanding universe described by a spatially-flat Friedmann-Lema\^itre-Robertson-Walker (FLRW) line element
\begin{eqnarray}
    ds^2 = g_{\mu\nu}dx^\mu dx^\nu = - a(\eta)^{2 \alpha} d\eta^2 + a(\eta)^2 \delta_{ij} dx^i dx^j\,, 
    \label{eq:FLRWmetric}
\end{eqnarray}
where $a(\eta)$ is the scale factor. The constant $\alpha$ can be chosen freely, defining the $\alpha$-time $\eta$, which becomes cosmic (conformal) time $t$ ($\tau$) for $\alpha = 0$ ($\alpha = 1$). Derivatives with respect to cosmic and $\alpha$-time will be denoted as $\dot f \equiv {df/dt}$ and $f' \equiv {df/ d\eta}$, respectively. Any set of equations of motion (EOM) implemented in the code can always be chosen to be solved in an arbitrary $\alpha$-time. %However, we might sometimes write EOM in this paper in cosmic time, just for clarity. 

The numerical implementation of any field sector in the code requires the fields to live in regular lattices of $N$ sites per dimension and side length $L$, where the minimum distance between two neighboring sites defines the \textit{lattice spacing} $\delta x \equiv L/N$. Such a lattice covers a range of discrete momenta, from a minimum infrared (IR) mode $k_{\rm IR} \equiv 2 \pi /L$ to a maximum ultraviolet (UV) scale $k_{\rm UV} = (N/2) k_{\rm IR}$ in each spatial dimension.

%%%%%%%%%%%%%%%%%%%%%%%%%%%%%%%%%%%%%%%%
\subsection{Scalar-gauge dynamics}
\label{subsec:scalargauge}

Since \CL{1.0} it is possible to solve the EOM of scalar and gauge fields in the expanding background determined by Eq.~\eqref{eq:FLRWmetric}. The code allows to simulate multiple copies of three kinds of scalar fields: real (singlet) scalars, complex scalars (that can be charged under U(1) gauge sectors), and scalar doublets (that can be charged under U(1) and/or SU(2) gauge sectors). We write these scalar fields in terms of real components, with normalizations chosen so that the components have canonically normalized kinetic terms:
\begin{eqnarray} \label{eq:ChargedScalars}
       \begin{array}{c|c|c}
           {\rm Singlet} & U(1){\rm-charged} & SU(2){\rm-charged~Doublet}
           \\\hline\vspace{-0.3cm} \, & \, &\\
           \phi \in \mathbb{R} &  \varphi \equiv {1\over\sqrt{2}}(\varphi_0 +i\varphi_1) & \Phi = \left(
           \begin{array}{c}
               \varphi^{(0)} \\ \varphi^{(1)}
           \end{array}
           \right) =
           {1\over\sqrt{2}}
           \left(
           \begin{array}{c}
               \varphi_0 +i\varphi_1 \vspace*{0.1cm}\\ \varphi_2 +i\varphi_3
           \end{array}
           \right)
\end{array}\,~~~~
\end{eqnarray}
One can also simulate both Abelian and non-Abelian SU(2) gauge fields, which we denote as $A_{\mu}$ and $C_{\mu} = C_{\mu}^aT_a$ respectively, with $T_a \equiv \sigma_a /2$ ($a=1,2,3$) the generators of the SU(2) group, and $\sigma_a$ the Pauli matrices. 

The most generic dynamics that can be solved by \CL{1.X} follow from the Lagrangian
\begin{align} 
    -\mathcal{L} = \frac{1}{2}\partial_{\mu} \phi \partial ^{\mu}\phi + (D_{\mu}^A \varphi)^{*}(D_A^{\mu} \varphi) +  (D_{\mu}\Phi )^{\dagger} (D^{\mu} \Phi) + \frac{1}{4} F_{\mu \nu} F^{\mu \nu} + \frac{1}{2}{\rm Tr}\{G_{\mu \nu}G^{\mu \nu}\} + V\,,
    \label{eq:lagrangian} 
\end{align}  
with $V = V(\phi,|\varphi|, |\Phi|)$ a potential that determines the interactions between all scalar fields. For simplicity, we have considered above a single species of each type of scalar, but we note that \CL{1.X} allows for any number of them, with different charges under the U(1) and SU(2) sectors.

This Lagrangian leads to the following EOM in an expanding background~\cite{Figueroa:2020rrl}
\begin{eqnarray}
    \phi'' - a^{-2(1 - \alpha)} {\vv\nabla}^{\,2} \hspace{-1mm}\phi + (3 - \alpha)\mathcal{H} {\phi'} &=& - a^{2 \alpha} V_{,\phi} \ , \label{eq:singletEOM} \\
    \varphi'' - a^{-2(1 - \alpha)} {\vv D}_{\hspace{-0.5mm}A}^{\,2}\varphi + (3 - \alpha) \mathcal{H}  {\varphi'} &=& - \frac{a^{2 \alpha}V_{,|\varphi|} }{2} \frac{\varphi}{|\varphi |} \ , \label{eq:higgsU1EOM}\\
    \Phi'' - a^{-2(1 - \alpha)} {\vv D}^{\,2}\Phi + (3 - \alpha) \mathcal{H}  {\Phi'} &=& - \frac{a^{2 \alpha} V_{,|\Phi|}}{2} \frac{\Phi}{|\Phi |} \ , \label{eq:higgsSU2EOM}
    \\
    \partial_0 F_{0i} - a^{-2(1 - \alpha )}\partial_j F_{ji} + (1 - \alpha) \mathcal{H} F_{0i} &=&
    a^{2 \alpha}J^A_i \ , \label{eq:U1EOM}
    \\[1mm]
    (\mathcal{D}_0 )_{a b} (G_{0i})^b - a^{-2(1 - \alpha )} ( \mathcal{D}_j )_{a b} (G_{ji} )^b + (1 - \alpha) \mathcal{H} (G_{0i} )_a &=& a^{2 \alpha}(J_i)_a \ , \label{eq:SU2EOM}
    \\[2mm]
    \partial_i F_{0i} &=& a^2J^A_0 \ , \label{eq:GaussU1}\\[1mm]
    (\mathcal{D}_i )_{a b} (G_{0i})^b &=& a^2(J_0)_a \ , \label{eq:GaussSU2}
\end{eqnarray}
where $\mathcal{H} \equiv \frac{a'}{a}$ is the Hubble rate in $\alpha$-time. We note that Eqs.~(\ref{eq:GaussU1})-(\ref{eq:GaussSU2}) represent the $\mathrm{U}(1)$ and $\mathrm{SU}(2)$ Gauss constraints, respectively, rather than EOM. The Abelian and non-Abelian currents are $J_A^\mu \equiv 2 g_AQ_A^{(\varphi)} \mathcal{I}m [ \varphi^{*} ( D_A^{\mu} \varphi )] + 2 g_AQ_A^{(\Phi)} \mathcal{I}m [ \Phi^\dag (D^{\mu} \Phi  )]$ and $J_a^\mu \equiv 2g_CQ_C\mathcal{I}m [ \Phi^{\dag} T_a( D^{\mu} \Phi )]$, whereas $(\mathcal{D}_{\nu}O)_a = (\mathcal{D}_{\nu})_{a b}O_b \equiv ( \delta_{a b}  \partial_{\nu} - f_{abc} C_{\nu}^c ) O_b$, , and the {\it covariant derivatives} and {\it field strength tensors} %associated to the gauge fields 
are defined by
\begin{eqnarray}
    D_{\mu}^{ A}  &\equiv &  \partial _{\mu} - i  g_A Q_AA_\mu \ , \hspace{3cm} F_{\mu \nu}\equiv  \partial_{\mu}  A_{\nu} - \partial_{\nu} A_{\mu} \ , \label{eq:AbCovDerivCont} \\
    D_{\mu} & \equiv  &
    \mathcal{I}D^{ A}_\mu
    - i g_C Q_C C_{\mu}^a \,T_a
    \ , \hspace{2.13cm}  G_{\mu \nu} \equiv \partial_{\mu} C_{\nu} - \partial_{\nu} C_{\mu} - i[C_\mu,C_\nu]\,,  \label{eq:CovDerivCont} 
\end{eqnarray}
with $Q_{A}$ and $Q_C$ the Abelian and non-Abelian charges of the scalar fields, $g_{A}$ and $g_C$ the Abelian and non-Abelian gauge couplings, and $\mathcal{I}$ the $N\times N$ identity matrix. For later convenience, we also define electric and magnetic fields of the Abelian and non-Abelian fields as (here $\epsilon_{ijk}$ is the Levi-Civita symbol in three dimensions, with $\epsilon_{123}=+1$)
\begin{align}
    \label{eq:ElectricMagneticDefs}
    E_i \equiv F_{0i} , \,\,\,\,\,\,\,\,  B_i \equiv \frac{1}{2} \epsilon_{i j k} F^{j k} , \,\,\,\,\,\,\,\,   E_i^a \equiv G_{0i}^a , \,\,\,\,\,\,\,\,  B_i^a \equiv \frac{1}{2} \epsilon_{i j k} G^{j k}_a \ ,
\end{align}
where $G_{\mu \nu}^a \equiv {\rm Tr}(2G_{\mu \nu} T_a)$, and we note that the electric field definitions depend on the $\alpha$-time $\eta$, as $F_{0i}$ and $G_{0i}$ are defined with respect to $\eta$, not $t$.

The evolution of the scale factor %in the case of \textit{self-consistent} expansion, {\it i.e.}~when it is sourced by the actual fields of a simulation, 
is determined simultaneously with the evolution of the fields through the Friedmann equations,
\begin{align}
    {a''\over a} &= \frac{a^{2 \alpha}}{6 m_\text{p}^2}\Big[ (2 \alpha - 1) \bar{\rho} - 3 \bar{ p} \Big]\,,  \label{eq:Friedmann1} \\
    \left({a'\over a}\right)^2 &= a^{2 \alpha} \frac{\bar {\rho}}{3 m_\text{p}^2}  \,,\label{eq:Friedmann2}
\end{align}
where $\bar\rho \equiv \langle \rho \rangle$ and $\bar p \equiv \langle p \rangle $ are volume averages of the energy and pressure densities. \CL{1.X} can simulate the field dynamics either using an {\it external} expanding background characterized by a constant equation of state $w$, or via \textit{self-consistent expansion} of the Universe sourced by the actual fields of the simulation. In the latter case, the code solves Eq.~\ref{eq:Friedmann1} using the {\it local} expression of the field's energy and pressure densities,
\begin{align}
    \rho &= {K}_{\phi} + {K}_{\varphi} + {K}_{\Phi} + {G}_{\phi} + {G}_{\varphi} + {G}_{\Phi} + {K}_{\mathrm{U}(1)} + {G}_{\mathrm{U}(1)} + {K}_{\mathrm{SU}(2)} + {G}_{\mathrm{SU}(2)} + {V} \ ,  \label{eq:rhoLocal}\\
    p &= {K}_{\phi} + {K}_{\varphi} + {K}_{\Phi} -{1\over3}({G}_{\phi} + {G}_{\varphi} + {G}_{\Phi}) + %+ \nonumber \\&& 
    {1\over3}({K}_{\mathrm{U}(1)} + {G}_{\mathrm{U}(1)}) + {1\over3}({K}_{\mathrm{SU}(2)} + {G}_{\mathrm{SU}(2)}) - {V} \ ,
\end{align}
where %with the different energy density contributions given by
\begin{eqnarray}
    &&\hspace*{-0.5cm}\begin{array}{l} 
        {K}_{\phi} = \frac{1}{2 a^{2\alpha} } {\phi'}^2 \vspace{0.1cm}\\
        {K}_{\varphi} = \frac{1}{a^{2\alpha} } (D_0^A \varphi)^*(D_0^A \varphi)
        \vspace{0.1cm}\\
        {K}_{\Phi} = \frac{1}{a^{2\alpha} } (D_0 \Phi )^\dag(D_0 \Phi)
        \vspace{0.1cm}\\
    \end{array}
    ;\hspace{0.2cm}
    \begin{array}{l}
        {G}_{\phi} = \frac{1}{2 a^2} \sum_i (\partial_i \phi)^2
        \vspace{0.1cm}\\
        {G}_{\varphi} = \frac{1}{a^2} \sum_i (D_i^A \varphi)^*(D_i^A \varphi)
        \vspace{0.1cm}\\
        {G}_{\Phi} = \frac{1}{a^2} \sum_i (D_i \Phi)^\dag(D_i \Phi)
        \vspace{0.1cm}\\
    \end{array}
    ;\hspace{0.2cm}
    \begin{array}{l}
        {K}_{\mathrm{U}(1)} = \frac{1}{2 a^{2 + 2 \alpha}}  \sum_{i} F_{0i}^2
        \vspace{0.1cm}\\
        {K}_{\mathrm{SU}(2)} = \frac{1}{2 a^{2 + 2 \alpha}}  \sum_{a,i} (G_{0i}^a)^2
        \vspace{0.1cm}\\
        {G}_{\mathrm{U}(1)} = \frac{1}{2 a^4}  \sum_{i,j<i} F_{ij}^2
        \vspace{0.1cm}\\
        {G}_{\mathrm{SU}(2)} = \frac{1}{2 a^4}  \sum_{a,i,j<i}  (G_{ij}^a)^2\vspace*{0.2cm}\\
    \end{array}
    \nonumber\\
    &&\hspace*{0.0cm}{\rm(Kinetic-Scalar)} ~~~~~~~~~~~~~~~{\rm(Gradient-Scalar)} ~~~~~~~~~~~~ {\rm (Electric ~\&~ Magnetic)} \hspace{0.0cm} \nonumber\\
    \label{eq:energy-contrib}
\end{eqnarray}
%
%where $K_x$ and $G_x$ are the kinetic and gradient energy densities of each field species [see e.g.~Eq.~(51) of Ref.~\cite{Figueroa:2020rrl} for their expressions]. 
All EOM, Eqs.~(\ref{eq:singletEOM})-(\ref{eq:SU2EOM}), together with Eq.~\ref{eq:Friedmann1}, are solved in simultaneously by symplectic algorithms (such as {\it leapfrog} or {\it verlet} methods) in \CL{1.X}, see Sects.~4 \& 5 of \ArtI~\cite{Figueroa:2020rrl}
for details. % on the adaptation of these algorithms to scalar-singlet and scalar-gauge theories. 
The Hubble constraint \eqref{eq:Friedmann2} is used to monitor the accuracy of the fields' evolution, though its level of preservation depends strongly on the numerical solver. On the other hand, 
as the discretization schemes used in $\mathcal{C}\mathcal{L}$ respect gauge invariance at the lattice level exactly, the $\mathrm{U}(1)$ and $\mathrm{SU}(2)$ Gauss constraints--- {\it c.f.}~Eqs.~(\ref{eq:GaussU1})-(\ref{eq:GaussSU2}) --- are consequently 
preserved up to machine precision throughout the evolution.

We note that, in practice, what the code actually solves are discretized versions of the above EOM, expressed in \textit{program variables}. These are a set of dimensionless field and spacetime variables, defined in terms of two energy scales: $f_*$, related to the amplitude of the dominant scalar field species, and $\omega_*$, related to the (inverse) time scale of the problem at hand. The default program variables are set by
\begin{align}
    \label{eq:FieldSpaceTimeNaturalVariables}
    d\tilde\eta &\equiv a^{- \alpha} \omega_* dt\ , \hspace{0.4cm}
    d\tilde x^i \equiv \omega_* dx^i\ , \\%\\\label{eq:FieldSpaceTimeNaturalVariablesII} && 
    \hspace{0.4cm}
    \tilde\phi &\equiv \frac{\phi}{f_*} \ , \hspace{0.4cm}
    \tilde\varphi \equiv \frac{\varphi}{f_*} \ , \hspace{0.4cm} \widetilde{\Phi} \equiv \frac{\Phi}{f_*} \ , \hspace{0.4cm}  \widetilde{A}_\mu \equiv \frac{A_\mu }{\omega_*} \ , \hspace{0.4cm} \widetilde C_{\mu}^a \equiv  \frac{C_{\mu}^a}{\omega_*} \ .
\end{align}

%%%%%%%%%%%%%%%%%%%%%%%%%%%%%%%%%%%%%%%%
\subsection{Gravitational waves}

Since \CL{ 1.2}, the code has the capability of simulating the production of gravitational waves (GWs) from both scalar and Abelian gauge fields.
GWs correspond to transverse and traceless spatial metric perturbations on top of the FLRW background, {\it i.e.}~$ds^2 = -a^{2\alpha}(\eta)d\eta^2 + a^2(\eta)(\delta_{ij} + h_{ij})dx^idx^j$, with $\partial_i h_{ij} = h_{ii} = 0$. Their EOM is found after linearizing the Einstein equations, as~\cite{Baeza-Ballesteros:2025tme,Caprini:2018mtu}
\begin{eqnarray}
    \label{eqn:EoMGWs}
    {h}_{ij}'' + (3-\alpha) \frac{a'}{a} h_{ij}' - a^{-2(1-\alpha)}\nabla^2 h_{ij} = \dfrac{2 }{m_\text{p}^2 a^{2(1-\alpha)}}\Pi_{ij}^{\text{TT}}\,, 
\end{eqnarray}
where $\Pi_{ij}$ %$ \equiv T_{ij} - p  a^2(t) (\delta_{ij} + h_{ij})$ 
is the anisotropic tensor of all fields that source GWs, %accounts the deviation from the perfect fluid form, 
and ${\rm TT}$ denotes transverse-traceless projection. In practice, we define an \textit{effective} anisotropic tensor as the part of $\Pi_{ij}$ that contains only non-zero TT contributions, {\it i.e.} $(\Pi_{ij}^{\rm eff})^{\rm TT} = \Pi_{ij}^{\rm TT}$.  For singlet scalar and scalar-gauge fields, one obtains~\cite{PhDthesisFigueroa}
\begin{eqnarray}
    {\Pi}_{i j}^{\mathrm{eff}}={\nabla}_i {\phi} {\nabla}_j {\phi} +2 \operatorname{Re}\left\{\left({D}_i {\varphi}\right)^*\left({D}_j {\varphi}\right)\right\}- (a^{-2 \alpha} {E}_i {E}_j+a^{-2} {B}_i {B}_j )\,. \label{eq:anitensor} 
\end{eqnarray}
\CL{1.2} solves Eq.~(\ref{eqn:EoMGWs}) sourced by~(\ref{eq:anitensor}), using the procedure proposed in Ref.~\cite{Garcia-Bellido:2007fiu}, which prevents having to go back-and-forth to Fourier space at every time step. In \CL{2.0} we further improve the procedure, introducing a variant of the method that reduces the memory requirements, namely allowing to store only 5 degrees of freedom ({\it dof})---  as opposed to 6 {\it dof} ---, in order to reconstruct the tensor perturbation $h_{ij}$, see Sect.~\ref{subsec:NewGW} for details.

%%%%%%%%%%%%%%%%%%%%%%%%%%%%%%%%%%%%%%%%
\subsection{Initial conditions}

Regarding initial conditions for scalar fields, since \CL{1.0} the user can initialize the fields with a homogeneous amplitude and time-derivative, $\bar{\phi}$ and $\bar{\phi'}$, and spatially-dependent fluctuations on top. These fluctuations are typically characterized by a power spectrum $\mathcal{P}_\phi(k)$, which characterizes the variance of the field as follows
\begin{eqnarray}
    \label{eq:SpectrumContinuum}
    \left\langle \delta \phi^2 \right\rangle = \int d\log k~\Delta_{\phi}(k)\,,~~\Delta_{\phi}(k) \equiv {k^3\over 2\pi^2}\mathcal{P}_{\phi}(k)\,,~~
    \left\langle {\delta \phi}_{\bf k}{\delta \phi}_{{\bf k}'} \right\rangle \equiv (2\pi)^3 \mathcal{P}_{\phi} (k)\delta(\bf{k}-\bf{k}')\,.
\end{eqnarray}
%
%An analogous expression applies for the conjugate momenta $\pi_{\phi} \equiv a^{3 - \alpha} \phi$. 
By default, \CL{1.X} imposes a power spectrum of scalar fluctuations mimicking quantum vacuum fluctuations, given by
\begin{eqnarray}
    \label{eq:QuantumFlucts}
    \mathcal{P}_{\phi} (k) \equiv 
    \frac{1}{2 a^2\omega_{k,\phi}}\,,~~~~ \omega_{k,\phi} \equiv \sqrt{k^2 + a^2m_{\phi}^2} \,,~~~~ m_{\phi}^2 \equiv \frac{\partial^2 V}{\partial \phi^2}\Big|_{\phi = \bar{\phi}} \ . 
\end{eqnarray}
More specifically, the fluctuations of scalar fields are set in Fourier space through the expressions
\begin{eqnarray}
    %\delta  \phi ({  \bf k}) &=& \frac{1}{\sqrt{2}} (|\delta  \phi^{(l)} ({  \bf k})|  e^{i \theta^{(l)} ({   \bf {k}}) } + |\delta  \phi^{(r)} ({   \bf {k}})| e^{i \theta^{(r)} ({   \bf {k}}) }   ) \label{eq:fpr_influct} \ , \\
    \delta  \phi ({  \bf k}) &=& \frac{1}{\sqrt{2}} \big[\delta  \phi^{(l)} ({  \bf k}) + \delta  \phi^{(r)} ({  \bf {k}})\big] \label{eq:fpr_influct} \ , \\
    \delta  {\phi}' ({   \bf {k}})
    %&=& {1\over a^{1-\alpha}}\left[\frac{i {\omega}_k}{\sqrt{2}}  \left(|\delta  \phi^{(l)} ({   \bf {k}})| e^{i \theta^{(l)} ({   \bf {k}}) } - |\delta  \phi^{(r)}  ({   \bf {k}})| e^{i \theta^{(r)} ({   \bf {k}}) }   \right)\right]  - \mathcal{H}  \delta  \phi ({   \bf {k}}) , \label{eq:fpr_influct2} 
    &=& {1\over a^{1-\alpha}}\left(\frac{i {\omega}_k}{\sqrt{2}}  \big[\delta  \phi^{(l)} ({   \bf {k}}) - \delta  \phi^{(r)}  ({   \bf {k}}) \big]\right)  - \mathcal{H}  \delta  \phi ({   \bf {k}})\ , 
    \label{eq:fpr_influct2} 
\end{eqnarray}
with both $\delta \phi^{(l,r)}$ complex fields ({\it left}- and {\it right}-movers), each with real and imaginary parts, $f_{\rm R} + if_{\rm I}$. The fields are initialized at each point ${\bf k}$ of the Fourier lattice by drawing independent random realizations of the real ($f_{\rm R}$) and imaginary ($f_{\rm I}$) parts from a Gaussian distribution with zero mean and variance $\sigma^2 \equiv {1\over2} \times \mathcal{P}_\phi(k)$, with $\mathcal{P}_\phi(k)$ from Eq.~(\ref{eq:QuantumFlucts}). 

On the other hand, gauge fields are initialized by imposing fluctuations in the charged scalar sectors such that the Gauss constraints~\eqref{eq:GaussU1}-\eqref{eq:GaussSU2} are fulfilled at the initial time, see Sect.~7.2 of \ArtI~\cite{Figueroa:2020rrl} for details. %The dynamical evolution of the system then preserve these constraints up to machine precision. 

%%%%%%%%%%%%%%%%%%%%%%%%%%%%%%%%%%%%%%%%%%%%%%%%%%%%%%%%%%%%%%%%%%%%%%%%%%%%%%%%
\section{\CL{2.0} -- New Physics}%\cosmolattice {\tt v2.0}}
\label{sec:NewPhysics}

%In the following we detail the new physics than \CL{2.0} can simulate.

\subsection{Scalar fields non-minimally coupled to gravity} \label{sec:NonMinCoup}

\CL 2.0 allows one to simulate the dynamics of a scalar field non-minimally coupled (NMC) to gravity, characterized by the following action
\begin{align}
    {\cal S}_{\rm NMC} &=
    \int d^{4}x \sqrt{-g} \left( \frac{1}{2}m_\text{p}^2R - \frac{1}{2}\xi R \phi^{2} -\frac{1}{2} g^{\mu\nu}\partial_{\mu}\phi\partial_{\nu}\phi - V(\phi,{\{\varphi_{\rm m}\}})
    %+ \mathcal{L}_{\rm m}[{\{\varphi_{\rm m}\}}]
    \right) \,,
    \label{eq:action}
\end{align}
where $R$ is the Ricci scalar, $g$ the determinant of the spacetime metric $g_{\mu\nu}$, and ${\{\varphi_{\rm m}\}}$ represents (minimally coupled) {\it matter} sectors. % are evaluated only at the FLRW background level, {\it c.f.}~Eq.~(\ref{eq:FLRWmetric}). 
The dimensionless parameter $\xi$ measures the strength of the {\it non-minimal} coupling. In the absence of gravitational perturbations, {\it i.e.}~evaluating $g_{\mu\nu}$, and hence also $R$ and $\sqrt{-g}$ on the FLRW background level, {\it c.f.}~Eq.~(\ref{eq:FLRWmetric}), %and using the metric in Eq.~\eqref{eq:FLRWmetric}, 
the EOM for the NMC field reads
\begin{align}
    \phi'' +(3-\alpha) \frac{a'}{a}\phi' - a^{-2(1-\alpha)}\nabla^2\phi  + a^{2\alpha} \left(\xi \bar R \phi + \frac{\partial V}{\partial \phi}\right) =0 \,,
    \label{eq:eomNMC}
\end{align}
where the (background) Ricci scalar is dictated by the FLRW metric as
\begin{equation}
    \bar R = \frac{6}{a^{2\alpha}} \left[\frac{a''}{a} +(1-\alpha)\left( \frac{a'}{a}\right)^{2}\right] \,.
    \label{eq:cosmic_R}
\end{equation}
To write an evolution equation for the scale factor $a(t)$, one first traces the Einstein equations and takes volume averages, %and using the trace of the energy-momentum tensor of the NMC field
%and obtain~\cite{Figueroa:2021iwm,Figueroa:2024asq}%, obtain
%\begin{align}
%  R = -\frac{1}{m_\text{p}^2}g^{\mu\nu}\left( T^\phi_{\mu\nu} +  T^{\rm m}_{\mu\nu}\right) &= -\frac{1}{m_\text{p}^2}\left( T_\phi +  T_{\rm m}\right) \,,\nonumber \\
%  &= -\frac{1}{m_\text{p}^2}\left(\left(6\xi -1\right) \left(\partial^{\mu}\phi\partial_{\mu}\phi  + \xi R\phi^2 \right) + 6\xi\phi \frac{\partial V}{\partial \phi} - 4V + T_m\right) \,.
%  \label{eq:EFEtr}
%\end{align}
leading to an expression for the background Ricci scalar~\cite{Figueroa:2021iwm,Figueroa:2024asq}
\begin{eqnarray}
    \bar R = \frac{F(\phi)}{m_\text{p}^{2}}\Big[\left(1-6\xi \right) \langle\partial^{\mu}\phi\partial_{\mu}\phi\rangle  + 4 \langle V\rangle- 6\xi\langle \phi V_{,\phi}\rangle-\langle T_{\rm m} \rangle \Big]\,,\label{eq:eomR}\\
    {\rm where}~~~ F(\phi) \equiv \frac{1}{1 + \left(6\xi -1\right)\xi \langle\phi^2\rangle /m_\text{p}^2 } \,,\hspace{1.5cm} 
\end{eqnarray}
with $T_m$ the trace of the energy momentum-tensor of the minimally-coupled sectors. Eq.~\eqref{eq:cosmic_R} then leads to a differential equation for the scale factor,
\begin{align}
   \frac{a''}{a} +(1-\alpha)\left( \frac{a'}{a}\right)^{2} =
   \frac{a^{2\alpha}F(\phi)}{6m_\text{p}^{2}}\Big[\left(1-6\xi \right) \langle\partial^{\mu}\phi\partial_{\mu}\phi\rangle  + 4 \langle V\rangle- 6\xi\langle \phi V_{,\phi}\rangle-\langle T_{\rm m} \rangle \Big]\,. \label{eq:piadot}
\end{align}
\CL{2.0} solves the above equation simultaneously with Eq.~(\ref{eq:eomNMC}), the EOM of the NMC field, and the EOM of the minimally-coupled sectors, using non-symplectic integrators such as the Runge-Kutta integrators discussed in Sect.~\ref{subsec:RK}.

%%%%%%%%%%%%%%%%%%%%%%%%%%%%%%%%%%%%%%%%
\subsection{ALP-gauge interactions} \label{sec:AxionInter}

\CL{2.0} also incorporates a new module for axion-like particles (ALPs) coupled to Abelian gauge fields that simulates the dynamics following from an action like
\begin{equation}
    { S}_{\rm ALP} = -\int {\rm d}x^4 \sqrt{-g}\left\lbrace \frac{1}{2}\partial_\mu \phi\partial^\mu\phi+V(\phi) - \frac{1}{4}F_{\mu\nu}F^{\mu\nu} + \frac{\alpha_{\Lambda}}{4}\frac{\phi}{m_\text{p}} F_{\mu\nu}\tilde F^{\mu\nu} \right\rbrace \ ,  
\end{equation}
where $\alpha_{\Lambda}$ is a dimensionless axion-gauge coupling constant. In the absence of gravitational perturbations, {\it i.e.}~restricting the metric to the FLRW background solution~(\ref{eq:FLRWmetric}), the variation of %$S_{\rm tot} = S_{\rm g} + S_{\rm m}$ with $S_{\rm g} \equiv \int {\rm d}x^4 \sqrt{-g}\,\frac{1}{2}m_\text{p}^2 R$ the standard Hilbert-Einstein action, 
${S}_{\rm ALP}$, together with the Friedmann equations, produces the following set of EOM (which we write here, for simplicity, in cosmic time)
\begin{eqnarray}\label{eq:AxionInfEOM}
    \left.
    \begin{array}{rcl}
        \ddot\phi &=& -3H\dot\phi+\frac{1}{a^2}\vec\nabla^2\phi-V_{,\phi}+\frac{\alpha_\Lambda}{a^3 m_\text{p}}\vec{E}\cdot\vec{B}\,,\label{eqn:eom1}\vspace*{1mm}\\
        \dot{\vec{E}} &=& -H\vec{E}-\frac{1}{a^2}\vec{\nabla}\times\vec{B}-\frac{\alpha_\Lambda}{a m_\text{p}}\Big(\dot\phi\vec{B}-\vec{\nabla}\phi\times\vec{E}\Big),\label{eqn:eom2}\vspace*{2mm}\\
        \ddot{a} &=&
        -\frac{a}{3m_\text{p}^2}\big\langle 2\rho_{\rm K}-\rho_{\rm V}+\rho_{\rm EM} \big\rangle\,, 
        \label{eqn:eom3}\vspace*{2mm}\\
        \vec{\nabla}\cdot\vec{E} \,\,&=& -\frac{\alpha_{\Lambda}}{a m_\text{p}}\vec{\nabla}\phi\cdot\vec{B}\,,\hspace*{3cm}{\rm\tt [Gauss~law]}\label{eq:Gauss}\vspace*{2mm}\\
        H^2\, &=& ~\frac{1}{3m_\text{p}^2}\big\langle \rho_{\rm K}+\rho_{\rm G}+\rho_{\rm V}+\rho_{\rm EM}\big\rangle\,,{\rm\tt ~~[Hubble~law]}
        \label{eq:Hubble}
    \end{array}\qquad\right\rbrace
\end{eqnarray}
%
%with $\vec B \equiv \vec\nabla \times \vec A$ the magnetic field, $\vec E \equiv \partial_t{\vec A}$ the electric field (in the temporal gauge $A_0 = 0$), and 
where the electromagnetic and inflaton's kinetic, potential and gradient energy densities are given by $\rho_{\rm EM} \equiv \frac{1}{2a^4}\langle a^2\vec E^2+\vec B^2\rangle$, $\rho_{\rm K} \equiv \frac{1}{2}\langle\dot{\phi}^2\rangle$, $\rho_{\rm V} \equiv \langle V \rangle$, and $\rho_{\rm G} \equiv \frac{1}{2a^2}\langle(\vec\nabla\phi)^2\rangle$, respectively, and $\langle ... \rangle$ denotes volume averaging. While the first three equations describe the system dynamics, the last two represent constraint equations. 

\CL{2.0} can solve the system of equations given by the first three equations in Eq.~(\ref{eq:AxionInfEOM}) recast in $\alpha$-time with the Runge-Kutta solvers introduced in Sect.~\ref{subsec:RK}. Throughout evolution, it keeps track of the Gauss and Hubble constrains. This module has already been successfully used {\it e.g.}~to understand the back-reaction regime of axion-inflation scenarios, see Refs.~\cite{Figueroa:2023oxc,Figueroa:2024rkr,Lizarraga:2025aiw}. The code can also output separately the spectra of the two chiralities of the gauge field. For more infomation on the ALP-gauge module of \CL{2.0}, see the $\texttt{Axion-Gauge Interactions}$ tab of the \href{http://www.cosmolattice.com}{\color{blue} \cosmolattice website}\,.

%%%%%%%%%%%%%%%%%%%%%%%%%%%%%%%%%%%%%%%%
\subsection{Cosmic defects}

Since \CL{1.0}, one can, in principle, simulate the formation and evolution of {\it global} cosmic defects, such as domain walls, strings, monopoles and textures, or even of {\it local} cosmic strings. This is because the EOM that evolve these objects are a subset of the Eqs.~\eqref{eq:singletEOM}-\eqref{eq:SU2EOM} that \CL{1.X} can solve. The manner by which \CL{1.X} initializes fields by default, setting up quantum-like fluctuations [{\it c.f.}~Eq.~(\ref{eq:SpectrumContinuum})] over homogeneous modes, can be used in practice to study the initial formation stage of the defects in certain cases. However, one is often interested in studying the long-term evolution of the defects during the {\it scaling} regime. %, or even their eventual annihilation (which is essential {\it e.g.}~for domain wall scenarios). 

Furthermore, simulations of topological defects are particularly challenging due to the need to resolve their cores at small scales ({\it e.g.}~width of a local string), while also capturing a sufficiently large number of defects in the lattice volume. This is because in an expanding universe, the separation between the core width of the defect (fixed by the inverse mass scales) and the average distance between the defects (growing approximately with the Hubble radius in scaling), increases with time. Together, this leads to the necessity of lattices with an extremely large number of sites per dimension, $N \sim 10^3-10^4$. 

Due to the problems just highlighted, it is convenient to envisage more appropriate initialization procedures than just the use of random vacuum fluctuations. The new cosmic defect module implemented in \CL 2.0 tackles precisely the aforementioned issues. To begin with, the code uses a set of artificial procedures to initially bring a network of defects very close to the scaling regime, before standard physical evolution is applied. This follows the ideas from Refs.~\cite{Hindmarsh:2017qff,Hindmarsh:2019csc,Hindmarsh:2021vih,Correia:2024cpk}. The code allows one to consider first a random initialization that mimics the symmetry breaking process  responsible for the defects. For example, \CL{2.0} allows the user to create scalar field fluctuations with power spectrum
\begin{equation}\label{eq:initialPSstrings}
    \Delta_{\phi_i}(k) = \frac{k^3 v^2 \ell_\text{str}^3}{\sqrt{2\pi}}\text{exp}\left(-\frac{1}{2}k^2\ell^2_\text{str}\right)\,,
\end{equation}
so that the system lies on the broken phase with the field configuration ensured to obey $\sum_a\langle \tilde{\phi}_a^2 \rangle =1$, but having randomly distributed phases. The conjugate momentum is initialized to zero. The expression in Eq.~(\ref{eq:initialPSstrings}) depends on a length scale $\ell_\text{str}$, which is a tuneable parameter that controls the initial density of the will-be network of cosmic strings. Other similar methods mimicking symmetry breaking are also available since \CL{2.0}. 

Regardless of the random process used for the initial conditions, the resulting configuration can be then evolved with a {\it dissipative} method, using a {\it diffusion} equation~\cite{Hindmarsh:2019csc,Hindmarsh:2021vih,Correia:2024cpk}. This dissipates the excess energy from the symmetry breaking process more efficiently than canonical evolution would do in an expanding background, hence rapidly leading to a network of well-formed and localized defect configurations. After this dissipative phase, the resulting network is expected to arrive much faster at the scaling regime via canonical evolution. The diffusion equations are presented in Eqs.~(191), (214), and (231) of \ArtII~\cite{Baeza-Ballesteros:2025tme} for global strings, local strings, and domain walls, respectively. 

Due to the natural loss of resolution of defect cores during physical evolution, \CL{2.0} also complements the above dissipative process with resolution-preserving methods for cosmic strings. We use the so-called {\it fattening} technique~\cite{Press:1989yh,Bevis:2006mj,Moore:2001px}, by which the comoving width of the defects is artificially maintained constant, ensuring the resolution of the defect width until the end of the simulation. Alternatively, the {\it extra-fattening} method is also available in the code, by which previously diffused fields are initially evolved with a set of equations that allow the comoving core radius to grow proportionally to the scale factor. Afterwards, this is followed by a phase of standard evolution in which the comoving width of the strings decreases, so that the string resolution at the end of the simulation is the same as it was at the beginning. The extra-fattening phase can be considered as part of the preparation of the initial condition, so that the strings are sufficiently well resolved on the lattice when the scaling regime is finally reached during physical evolution. The form of the field EOM during these special phases is presented for global and local cosmic strings in Eqs.~(194) and (216), respectively, in \ArtII~\cite{Baeza-Ballesteros:2025tme}, where a single parameter $s$ controls the regime: extra-fattening ($s = -1$), fattening ($s = 0$), and physical evolution ($s = 1$). 

Finally, \CL{2.0} also incorporates new observables especially convenient for the study of defects, such as {\it e.g.}~total string length estimated from the number of plaquettes with non-zero {\it winding} number, the area parameter of domain walls, or weighted energy components of the defects. For further information on the defect capabilities of \CL{2.0}, see the {\tt Cosmic Defects} tab of the online manual in \href{http://www.cosmolattice.com}{\color{blue} \cosmolattice website}; for the defect observables in particular, see the {\it Defect-specific observables} sub-tab there.

%%%%%%%%%%%%%%%%%%%%%%%%%%%%%%%%%%%%%%%%
\subsection{Scalar dynamics in $1+1$ and $2+1$ dimensions}

Simulating scalar fields on lattices with $d \neq 3$ spatial dimensions may be useful for two main reasons. First, one may wish to simulate a field theory intrinsically defined in $d \neq 3$ dimensions, with scalar fields propagating, {\it e.g.}, in a $(2+1)$-dimensional FLRW spacetime, $ds^2 = a^2(\eta)(-d\eta^2 + \delta_{ij}dx_i dx_j)$, with $i,j=1,2$. In this case, the corresponding EOM in $(2+1)$ dimensions can be discretized and solved using suitable evolution algorithms, much the same way as for theories in $(3+1)$ dimensions. 

Alternatively, one may wish to simulate a field theory defined in $d=3$ spatial dimensions on a lower-dimensional lattice, thereby approximating its three-dimensional dynamics at a reduced computational cost. This approach is justified, provided that the system is statistically isotropic, both in its initial conditions and throughout its evolution. This is typically the case for scalar-field interactions, as they can not create locally (neither globally) a preferred spatial direction (unless that is already imposed as an initial condition in the field gradients). Reducing the dimensionality from $d=3$ to $d=2$, can be therefore highly advantageous, accelerating simulations by a factor $\sim N$, with $N \approx 10^2-10^3$ in most applications. This enables investigations requiring very long evolution times, very large lattices, or extensive scans over model parameters, see {\it e.g.}~\cite{Antusch:2020iyq,Antusch:2021aiw,Antusch:2025ewc}. The ability of lower-dimensional simulations to accurately reproduce three-dimensional dynamics must be assessed anyways on a case-by-case basis.

\CL{2.0} incorporates the second circumstance, namely simulating $(3+1)$-dimensional physics on a reduced $(d+1)$-dimensional lattice, with $d=2$, or $d=1$. For such a purpose, the code allows to solve the singlet scalar field EOM in both $(1+1)$ and $(2+1)$ dimensions. More specifically, the code solves EOM of the form given in Eq.~\eqref{eq:singletEOM}
and Eq.~\eqref{eq:Friedmann1}, but on a $1-$ or $2-$dimensional spatial slice respectively. The discrete Laplacian operator in Eq.~\eqref{eq:singletEOM} is then computed summing over the $d$-spatial directions, {\it i.e.}~$\nabla^2 \phi = \sum_{i=1}^{d} \partial^2 \phi / \partial x_i^2$, with $d = 1$ or $2$. A similar change applies to the gradient component of the scalar field's energy density, ${G}_{\phi} = \frac{1}{2 a^2} \sum_{i=1}^d (\partial\phi / \partial x_i)^2$, {\it c.f.}~Eq.~\eqref{eq:energy-contrib}. By additionally modifying appropriately the initial spectrum of fluctuations, these simulations mimic the three-dimensional dynamics, see Sect.~7.1 of \ArtII~\cite{Baeza-Ballesteros:2025tme} for more extensive discussions.

%%%%%%%%%%%%%%%%%%%%%%%%%%%%%%%%%%%%%%%%%%%%%%%%%%%%%%%%%%%%%%%%%%%%%%%%%%%%%%%%
\section{\CL{2.0} -- New Features}
\label{sec:NewFeatures}

%%%%%%%%%%%%%%%%%%%%%%%%%%%%%%%%%%%%%%%%
\subsection{Non-symplectic evolvers}
\label{subsec:RK}

The EOM of interacting fields in an expanding universe can be schematically written in a common form, independently of the field content. To see this, let us denote a set of fields as $\lbrace f_{j}\rbrace$, and their conjugate momenta as $\lbrace \pi_j \rbrace$, with the index $j$ labeling the different $dof$. The form of their EOM can be encapsulated by
\begin{eqnarray}\label{eq:SchemeContVirgin1}
    \pi_a(\eta) &=& a'(\eta)\,,\\
    \label{eq:SchemeContVirgin2}
    \pi_a'(\eta) &=& \mathcal{K}_a[a(\eta), E_V(\eta), E_K(\eta), E_G(\eta)]\,,\\
    \label{eq:SchemeContVirgin3}
    \pi_i({\bf x},\eta) &=& \mathcal{D}_i[f_i'({\bf x},\eta),a(\eta),\pi_a(\eta);\lbrace f_{j}({\bf x},\eta) \rbrace, \lbrace f'_{j\neq i}({\bf x},\eta) \rbrace]\,,\\
    \label{eq:SchemeContVirgin4}
    \pi_i'({\bf x},\eta) &=& \mathcal{K}_i[f_i({\bf x},\eta),\pi_i({\bf x},\eta),a(\eta),\pi_a(\eta);\lbrace f_{j\neq i}({\bf x},\eta) \rbrace, \lbrace \pi_{j\neq i}({\bf x},\eta) \rbrace]\,,
\end{eqnarray}
where for each field, indicated by $i$, the \textit{drift}, $\mathcal{D}_i[...]$, defines the different conjugate momenta, and the \textit{kernel} or \textit{kick}, $\mathcal{K}_i[...]$, determines their interactions with other $dof$ (including possibly themselves). In the case of self-consistent expansion of the Universe, the scale factor is considered as a homogeneous $dof$, with Eq.~\eqref{eq:SchemeContVirgin2} corresponding to the Friedmann equation \eqref{eq:Friedmann1}. 

In the case of \textit{canonical} field interactions, as the scalar-gauge sectors or the GW dynamics considered in \CL{1.X}, the kernels do not depend on the conjugate momenta\footnote{This can be achieved for all EOM implemented in \CL{1.X}, as long as the conjugate momenta are appropriately defined, see \ArtI~\cite{Figueroa:2020rrl}.}, {\it i.e.}~$\mathcal{K}_i({\bf x},\eta) \equiv \mathcal{K}_i[\lbrace f_{j}({\bf x},\eta)\rbrace,a(\eta)]$. On the contrary, for non-canonical interactions, the kernels depend explicitly on the conjugate momentum, {\it i.e.}~$\partial \mathcal{K}_i / \partial\pi_i \neq 0$. This is the case, {\it e.g.}~of non-minimally coupled-to-gravity scalar fields (Sect.~\ref{sec:NonMinCoup}) or axion-gauge interactions (Sect.~\ref{sec:AxionInter}), implemented in \CL{2.0}. 

\CL{1.X} implemented \textit{symplectic} algorithms adapted to solve the EOM of the scalar-gauge matter sectors, {\it c.f.}~Eqs.~\eqref{eq:singletEOM}-\eqref{eq:SU2EOM}, and the EOM for the GWs, {\it c.f.}~Eq.~\ref{eqn:EoMGWs}, Namely, \textit{staggered leapfrog} and \textit{velocity verlet} methods of accuracy $\mathcal{O}(\delta\eta^2)$, and {\it Yoshida} algorithms of accuracy $\mathcal{O}(\delta\eta^4)$-$\mathcal{O}(\delta\eta^{10})$. Non-canonical interactions, however, require non-symplectic integration methods, in order to maintain stability, see \ArtI~\cite{Figueroa:2020rrl} for discussion. \CL{2.0} incorporates for this purpose a family of Runge-Kutta (RK) algorithms. The code can use standard RK methods of $\mathcal{O}(d\eta ^2)$ (RK2) or $\mathcal{O}(d\eta ^4)$ ($RK4$). Both of these methods execute intermediate sub-steps that require {\it auxiliary fields} to store extra information at each step, requiring one (three) auxiliary field(s) per field $dof$ for RK2 (RK4). Furthermore, \CL{2.0} also implements {\it low-storage} RK methods~\cite{Carpenter1994Thirdorder2R,Carpenter1994Fourthorder2R,Bazavov:2021pik,Bazavov:2025dzo,Bazavov:2025exj}, which represent a refined version of the previous schemes, in which the number of auxiliary fields is reduced to only one / $dof$, maintaining the integration accuracy, but at the expense of introducing further intermediate sub-steps. In \CL{2.0} we have considered low-storage RK methods ($\mathrm{RK}n$-$s$) of $n$-th order and $s$ sub-steps, in particular $\mathrm{RK}3$-$3$, $\mathrm{RK}3$-$4$, and $\mathrm{RK}4$-$5$. 

We note that while the RK algorithms implemented in \CL{2.0} can also be used to solve the dynamics of canonical interactions [{\it e.g.}~Eqs.~(\ref{eq:singletEOM})-(\ref{eq:SU2EOM}) with Eq.~(\ref{eq:Friedmann1})], or of GWs [{\it c.f.}~Eq.~(\ref{eqn:EoMGWs})], compared to symplectic methods, they are not optimal for such cases. Non-symplectic integrators of the RK family represent, in general, a set of versatile methods with broad applicability, capable of handling non-Hamiltonian, dissipative, or stiff systems, such as {\it e.g.}~non-canonical interactions with conjugate momenta in the kernels. For example, these algorithms can be easily applied to solve the EOM of scalar fields with non-canonical kinetic terms, see Sect.~3.2 of \cite{Baeza-Ballesteros:2025tme}. Furthermore, a major extension upgrade of the field content of the code is planned for \CL{3.0}, where we will incorporate fluids~\cite{Figueroa:2026esg} (in isolation or coupled to scalar or gauge sectors), for which it is crucial to use RK methods. Besides, some of the RK methods can naturally accommodate adaptive time-stepping. This is the case of our RK3-4 for instance.

%%%%%%%%%%%%%%%%%%%%%%%%%%%%%%%%%%%%%%%%
\subsection{New algorithm for GWs}
\label{subsec:NewGW}

While the equation governing the dynamics of gravitational waves (GWs) is linear, {\it c.f.}~Eq.~\eqref{eqn:EoMGWs}, the transverse-traceless (TT) projection of its effective source~\eqref{eq:anitensor} corresponds to a non-local operation in position-space, which requires to be obtained at every time step. In Ref.~\cite{Garcia-Bellido:2007fiu} a workaround was proposed to overcome this computational problem: noting that obtaining the TT-part of a 2-rank tensor corresponds to a local linear operation in momentum space, one can solve a wave equation in real space for several unphysical {\it dofs}, $ u_{ij} $, with the $lhs$ equal to that of Eq.~\eqref{eqn:EoMGWs}, but the $rhs$ given by the effective source $\Pi_{ij}^{\rm eff}$ without TT-projection. Then, only when desired to obtain the physical degrees of freedom truly representing GWs, $h_{ij}$, one Fourier-transforms the  $u_{ij}$ functions, and from there builds via TT projection, for example, the energy density of the GW background (GWB) present in the simulation. This procedure, adopted in \CL{1.1} for scalar field sources and in \CL{1.2} for scalar-gauge (Abelian) sources, requires 6 $dof$, $\lbrace u_{11}, u_{12}, u_{13}, u_{22}, u_{23}, u_{33} \rbrace$, and prevents having to go back-and-forth to Fourier space in every evolution step; see Ref.~\cite{Garcia-Bellido:2007fiu} for details. 

In \CL{2.0} we have incorporated a variant of the method that improves the memory requirement, namely allowing to store only 5 {\it dof} %(as opposed to $6$) 
to reconstruct the physical GW perturbation $h_{ij}$. The procedure, based on the algorithm presented in \ArtII~\cite{Baeza-Ballesteros:2025tme}, consists in solving the EOM of some new auxiliary variables $\lbrace v_{ij} \rbrace$, which are already forced to be traceless, via the condition $v_{33}=-v_{11}-v_{22}$. In this case, the EOM for $\lbrace v_{11}, v_{12}, v_{13}, v_{22}, v_{23} \rbrace$ are
\begin{eqnarray}
    {v}_{ij}'' + (3-\alpha) \frac{a'}{a} v_{ij}' - a^{-2(1-\alpha)}\nabla^2 v_{ij} = \dfrac{2 }{m_\text{p}^2 a^{2(1-\alpha)}}\left[\Pi_{ij}^\text{eff}-\frac{1}{3}\delta_{ij}\Pi^\text{eff}\right]\,,
\end{eqnarray}
where $\Pi^{\rm eff} \equiv \sum_k\Pi^{\rm eff}_{kk}$ is the trace of the effective anisotropic stress tensor. The physical GWs represented by TT-{\it dof} are obtained through the relation $h_{ij} (\vec{k},t) = \Lambda_{ij,kl} ({\bf \hat{k}}) v_{k l} (\vec{k},t)$, with $\Lambda_{ij,kl} ({\bf \hat{k}})$ the standard local TT-projection operator~\cite{Caprini:2018mtu}, $\Lambda_{ij,lm}({\hat{\bf k}}) \equiv P_{il}({\hat{\bf k}})  P_{jm}({\hat{\bf k}}) - 0.5 P_{ij}({\hat{\bf k}}) P_{lm}({\hat{\bf k}})$, with $P_{ij}(\hat{\bf k}) \equiv \delta_{ij} - {\hat{ k}}_i {\hat{k}}_j$, and ${\hat{ k}}_i \equiv {k_i/ k}$. %Thanks to the fact that $P_{ij}$ is transverse ($P_{ij} \hat{k}_j = 0$) and  idempotent ($P_{ij}P_{jm} = P_{im}$), one can easily see that the TT conditions in Fourier space, $k_i\Pi_{ij}^{\rm TT}({\bf {k}},\eta)=\Pi_{ii}({\bf {k}},\eta)^{\rm TT} = 0$, are satisfied at any time. 
The code also obtains the spectrum of the energy density of the GWs present in a simulation and its power spectrum at any time, via discretization of the formula~\cite{Figueroa:2011ye}
\begin{eqnarray} 
    \rho_{\rm GW} (t) \equiv \int \frac{d \rho_{\rm GW}}{d \log k} d \log k\,,~~~~
    \frac{{\rm d} \rho_{\rm GW}}{{\rm d} \log k} \equiv \frac{m_\text{p}^2 k^3}{8 \pi^2 a^{2\alpha} V} \int \frac{{\rm d} \Omega_k}{4 \pi} h'_{ij} (\vec{k},t) h^{'*}_{ij} (\vec{k}, t) \,
\end{eqnarray}
with $V$ the (comoving) volume of the simulation. For further details on how to discretize these expressions, we point the interested reader to Sect.~8.2 of \ArtII~\cite{Baeza-Ballesteros:2025tme}, or to explore the Tab {\it Gravitational Waves} in the \href{http://www.cosmolattice.com}{\color{blue} \cosmolattice website}\,.

We note that the code actually outputs the normalized energy density as $\Omega_{\rm GW} \equiv \frac{1}{\rho_{\rm tot}} \frac{{\rm d} \rho_{\rm GW}}{{\rm d} \log k} $, where $\rho_{\rm tot}$ is the total energy density in the lattice, which coincides only with the critical energy density of the system, $\rho_{\rm tot} = \rho_c$, in the case of self-consistent expansion.

%%%%%%%%%%%%%%%%%%%%%%%%%%%%%%%%%%%%%%%%
\subsection{Arbitrary binning and unbinned spectra}

The ensemble average $\langle ... \rangle$ that serves the define the notion of power spectrum in the continuum, {\it c.f.}~Eq.~\eqref{eq:SpectrumContinuum}, 
is substituted on the lattice by a volume average as
\begin{eqnarray}
    \langle f^2 \rangle_V = \frac{\dx^3}{V}\sum_{\bf n} f^2({\bf n}) = \frac{1}{N^3}\sum_{\bf n} f^2({\bf n})~\,,
\end{eqnarray}
so that using the discrete Fourier transform $f({\bf n}) \equiv {1\over N^3}\sum_{\tilde {\bf n}} e^{+i{2\pi\over N} \tilde{\bf n}{\bf n}} f({\bf \tilde n})$, %$f(\tilde{\bf n}) \equiv \sum_{\bf n} e^{-i{2\pi\over N} \tilde{\bf n}{\bf n} }f({\bf n})$, 
we obtain
\begin{eqnarray}
    \langle f^2 \rangle_V = \frac{1}{N^6}\sum_{\tilde{\bf n}}\big|f(\tilde{\bf n})\big|^2 = \frac{1}{N^6}\sum_{l} \sum_{\tilde{\bf n}'\in R_l} \big|f(\tilde{\bf n}')\big|^2\,.
    		\label{eq:discretePSaux}
\end{eqnarray}
In the last expression we have decomposed the sum in two parts: $i)$ an angular direction $\sum_{\tilde{\bf n}'\in R_l}( ... )$, summing over all modes with moduli within spherical bin shells of radius $|\tilde{\bf n}'|\in R_l \equiv \big[l -\Delta\tilde{n}_l/2,l+\Delta\tilde{n}_l/2\big)$, with $l = 1,2,3,...$ counting the bins and $\Delta\tilde{n}_l$ characterizing the width of the $l$-th bin; and $ii)$ a radii summation $\sum_{l}$, summing over the actual bins defined by the chosen binning. This separation allows to define the notion of {\it isotropic modulus of a mode function} as $|f_{l}|^2 \equiv \sum_{|\tilde{\bf n}|\in R_l} \big|f(\tilde{\bf n})\big|^2$, so that $	\langle f^2 \rangle_V = \frac{1}{N^6}\sum_{l} |f_{l}|^2$. A natural choice for binning is to consider $\Delta\tilde n = 1$, so that $|f_{l}|^2 = |f_{|\tilde{\bf n}|}|^2 \equiv \sum_{\tilde{\bf n}
'\in R(\tilde{\bf n})} \big|f(\tilde{\bf n}')\big|^2$ represents a summation over the modes $\tilde{\bf n}'$ within an spherical shell $R(\tilde{\bf n}) \equiv \big[|\tilde{\bf n}| - 1/2, |\tilde{\bf n}| + 1/2\big)$, with $\tilde{\bf n}$ restricted only to $|\tilde{\bf n}| = 1, 2, 3, ...$. We refer to this as the {\it canonical binning}, implemented in \CL{1.X} (\CL{2.0}) as the only (default) option. 

While the canonical binning is typically a good choice, it is possible to conceive an arbitrary radial binning $R_l$ with bins of width $\Delta\tilde n \neq 1$. After all, the modes in the momentum-lattice, $\tilde{\bf n} = (\tilde n_1, \tilde n_2, \tilde n_3)$, with $\tilde n_i = -\frac{N}{2}+1, -\frac{N}{2}+2, ... ,-1,0,1, ... , \frac{N}{2} - 1, \frac{N}{2}$, $i  = 1,2,3$, contain momenta of moduli %$k = k(\tilde n) \equiv k_{\rm IR}|\tilde {\bf n}|$, where 
$\tilde n = |\tilde {\bf n}| \equiv (\tilde{n}_1^2+\tilde{n}_2^2+\tilde{n}_3^2)^{1/2}$, which do not need to be an integer number. \CL{2.0} allows the user to choose for regular bins an arbitrary width $\Delta \tilde n$, either bigger or larger than unity. We note that choosing $\Delta \tilde n < 1$ may lead however to an underrepresented statistical sampling of the modes of momentum close to $|{\bf k}| = k_{\rm IR}\tilde n$, when computing the power spectrum at that scale. %specially for small values of $\Delta \tilde n$, 
On the other hand, $\Delta \tilde n > 1$ may lead to a wrong representation of the spectrum at $|{\bf k}| = k_{\rm IR}\tilde n$, weighted by modes that sustain a spectral amplitude $\mathcal{P}_f(|{\bf k}'|)$ too different from $\mathcal{P}_f(|{\bf k}|)$. The appropriateness of choosing finer or thicker binning options than the canonical $\Delta\tilde n = 1$ case, must be therefore checked in a case by case basis. 

The finest binning one can think of consists on considering all different momentum moduli values in the reciprocal lattice, $\lbrace \tilde n \equiv (\tilde{n}_1^2+\tilde{n}_2^2+\tilde{n}_3^2)^{1/2} \rbrace$, each belonging to a different bin with their own (non-regular) width. \CL{2.0} allows for the computation of this {\it unbinned} spectrum, $\mathcal{P}_f(\tilde n)$, which takes as many values as the number of momentum moduli available in the Fourier lattice. As such number grows as $\sim N^2$ with the size of the lattice, \CL{2.0} only allows the unbinned spectrum to be saved in HDF5 format. We note that the statistical sampling in the computation of the unbinned power spectrum can be very underrepresented, specially in the IR scales of the lattice, where the numerical spectrum may oscillate wildly above and below the theoretical expectation ue to the low number of realizations (this is similar to the so-called {\it cosmic variance} effect in the low multipoles of the angular power spectrum of the cosmic microwave background temperature fluctuations). 

%DGF: In more UV scales, the statistical sampling approaches better the theoretical expectation. %, the better the closer we approach the Nyquist frequency $\tilde n = N/2$ ({\color{red} this really true?}). 

%%%%%%%%%%%%%%%%%%%%%%%%%%%%%%%%%%%%%%%%
\subsection{External initial spectra}

In \CL{2.0}, a field can be now initialized from an arbitrary power spectrum $\Delta_{f}(k)\equiv k^3\mathcal{P}_{f}(k)/(2\pi^2)$, where $\mathcal{P}_{f}(k)$ is the variance of its Fourier counterpart ${f}({\bf k})$, {\it c.f.}~Eq.~(\ref{eq:SpectrumContinuum}). While in \CL{1.X}, for scalar fields, such spectrum is enforced to reproduce quantum vacuum fluctuations, as in Eq.~(\ref{eq:QuantumFlucts}), in \CL{2.0} other choices are possible. For example, fluctuations may first be evolved linearly on a one-dimensional grid of momentum magnitudes, say with an external code (like {\tt Mathematica} or {\tt Phyton}), and the resulting linear spectrum can then be sourced into \CL{2.0}, say shortly before the onset of non-linear dynamics developed on the lattice. Alternatively, a theoretical prediction of $\mathcal{P}_{f}(k)$, other than quantum fluctuations, can be also used.  

In \CL{2.0} the Fourier field amplitudes are drawn at each site $\tilde{\bf n}$ of the Fourier lattice, from Gaussian distributions with vanishing mean and variances fixed by the prescribed spectrum. % $\Delta_{f}(k) \equiv k^3\mathcal{P}_{f}(k)/(2\pi^2)$. 
The lattice counterpart of the continuum power spectrum is defined as~\cite{Baeza-Ballesteros:2025tme}
\begin{eqnarray}\label{eq:discretePST2}
    \Delta_{f}(k(|{\bf \tilde{n}}|))
    \equiv
    \frac{k(\tilde{\bf n})}{2\pi}\frac{\delta x}{N^5}
    \#_{R(\tilde{\bf n})}
    \left\langle\left|f(\tilde{\bf n})\right|^2\right\rangle_{R(\tilde{\bf n})}
    = \frac{k^3(\tilde{\bf n})}{2\pi^2}\,
    \Upsilon_{|\tilde{\bf n}|}
    \left(\frac{\delta x}{N}\right)^3
    \left\langle\left|f(\tilde{\bf n})\right|^2\right\rangle_{R(\tilde{\bf n})}\,.
\end{eqnarray}
where $\langle\cdots\rangle_{R(\tilde{\bf n})}$ denotes angular averaging over spherical shells of width $\Delta \tilde n$, {\it i.e.}~$|\tilde{\bf n}'|\in[|\tilde{\bf n}|,|\tilde{\bf n}|+\Delta\tilde n)$, $\#_{R(\tilde{\bf n})}$ is the number of lattice sites in that shell, and we have introduced 
\begin{eqnarray}\label{eq:UpsilonII}
    \Upsilon_{|\tilde{\bf n}|}
    \equiv
    \frac{\#_{R(\tilde{\bf n})}}{4\pi|\tilde{\bf n}|^2}\,,
\end{eqnarray}
to distinguish between {\tt Type-I} spectra, for which the exact multiplicity is used in the code and hence $\Upsilon_{|\tilde{\bf n}|}\neq1$, and {\tt Type-II} spectra, for which $\#_{R(\tilde{\bf n})}=4\pi|\tilde{\bf n}|^2$ is used, so that $\Upsilon_{|\tilde{\bf n}|}=1$ drops from Eq.~(\ref{eq:discretePST2}). 

To introduce arbitrary power spectra as initial conditions in \CL{2.0}, we parametrize first the spectra of both field amplitudes and derivatives as~\cite{Baeza-Ballesteros:2025tme}
\begin{eqnarray}
    \mathcal{P}_{f}(k) \equiv \frac{1}{a^3} \mathcal{F}(k/a)\,,~;~~~\mathcal{P}_{{\tt f}'}(k) = \frac{a^{2\alpha}}{a^3} \mathcal{G}(k/a) \;,
\end{eqnarray}
with $\mathcal{F}, \mathcal{G}$ characterizing the spectral shapes, such that $\langle {f}^2 \rangle = \int {\rm d} \log k \; \frac{(k/a)^3}{2\pi^2} \mathcal{F}(k/a)$ and $\langle {f}'^2 \rangle = \int {\rm d} \log k \; a^{2\alpha} \frac{(k/a)^3}{2\pi^2} \mathcal{G}(k/a)$. 
Identifying the shell-angular average with a statistical average over Gaussian realizations, leads to identify the variances of the field and of its time derivative as
\begin{eqnarray}\label{eq:varFandDFgeneral}
    \left\langle|f(\tilde{\bf n})|^2\right\rangle_{\rm stat}
    \equiv
    \frac{1}{\Upsilon_{|\tilde{\bf n}|}}
    \left(\frac{N}{\delta x}\right)^3
    \frac{{\mathcal F}(\kappa/a)}{a^3}\,~;~~~
    \left\langle|f'(\tilde{\bf n})|^2\right\rangle_{\rm stat}
    \equiv
    \frac{1}{\Upsilon_{|\tilde{\bf n}|}}
    \left(\frac{N}{\delta x}\right)^3
    \frac{{\mathcal G}(\kappa/a)}{a^{3-2\alpha}}\,,
\end{eqnarray}
In practice, \CL{2.0} initializes $f(\tilde{\bf n})= R(\tilde{\bf n})+ I(\tilde{\bf n})$ and $f'(\tilde{\bf n})= R'(\tilde{\bf n})+ I'(\tilde{\bf n})$ at every Fourier-lattice site $\tilde{\bf n}$, with the real and imaginary components sampled independently from Gaussian distributions with zero-mean and variance given by $1/2$ times the corresponding variances in Eq.~\eqref{eq:varFandDFgeneral}.

%%%%%%%%%%%%%%%%%%%%%%%%%%%%%%%%%%%%%%%%
\subsection{Improved snapshots}

\CL{1.X} is capable of printing three-dimensional snapshots of the different energy components in HDF5 format, namely the volume-averages of the different contributions to \eqref{eq:rhoLocal}. In \CL 2.0, we have added the capability of printing also snapshots of all scalar singlet amplitudes $\phi$, and the absolute values of a complex scalar $|\varphi|$. Furthermore, in order to save memory in cases with a large number of lattice points, we have also implemented the possibility of printing any subset of the lattice of arbitrary dimension $N_x \times N_y \times N_z$ (with $N_{x,y,z} < N$), as well as to regularly skip a certain number of points.

%%%%%%%%%%%%%%%%%%%%%%%%%%%%%%%%%%%%%%%%
\subsection{Single precision}

\CL{1.X} runs by default with double precision. In this case, the minimum RAM usage is given by the following expression,
\be {\rm RAM}_{\rm min} = 8 \cdot 2 N_{\rm dof} \left(\frac{N}{1024}\right)^3 {\rm GB} \ , \label{eq:RamMin} \ee
where the factor $8$ corresponds to the size of a double in bytes, $N_{\rm dof}$ is the total number of field components in the system (1 for scalar singlets, 2 for complex scalars, 4 for scalar doublets, 3 for Abelian gauge fields, 8 for non-Abelian gauge fields, and 5 for GWs), and the factor $2$ accounts for the conjugate momenta of each field component. Additional memory is required to store the ghost cells when running a distributed simulation with MPI, so this must be understood as a minimum threshold. Also, an additional temporary $\textit{dof}$ may be needed to measure certain observables, such as the PS of magnetic fields.

In certain situations, however, one might require to run simulations with a very large number of points, say $N = 2^{12}-2^{14}$, for which the memory requirement overcomes the maximum memory available by the HPC resources of the user. In that regard, \CL{2.0} supports simulations with single precision, which makes it possible %, given $N$ and $N_{\rm dof}$ 
to reduce the RAM requirement by half (as a single-precision \texttt{float} is represented in four bytes). %Running with single precision is very simple: one just need to compile with an additional cmake flag and change one single line in the model file. 
However, the accuracy of single precision simulations must be checked in a case by case basis, as the roundoff error of single-precision arithmetric lies on the order of $\mathcal{O}(10^{-7})$, whereas double precision offers $\mathcal{O}(10^{-16})$.

%%%%%%%%%%%%%%%%%%%%%%%%%%%%%%%%%%%%%%%%%%%%%%%%%%%%%%%%%%%%%%%%%%%%%%%%%%%%%%%%
\section{\CL{2.0} -- Performance Improvements}
\label{sec:NewCapabilities}

Before we comment on the improved performances of \CL{2.0}, we highlight how the use of the new code compares to \CL{1.X}. From the point of view of a standard user, using \CL 2.0 is actually very similar to previous versions. As before, the user has to edit two files: $i)$ the \textit{model file} (e.g.~\texttt{lphi4.h}), which contains the details of the scalar-field content of the model to be simulated, and $ii)$ an \textit{input} file (e.g.~\texttt{lphi4.in}), which specifies the model parameters to be passed to the code. A typical workflow to compile and run a model would be:\\[-2mm]
%
%\begin{shell-sessioncode}
%cmake -DMODEL=lphi4 ../
%make cosmolattice
%./lphi4 input=lphi4.in
%\end{shell-sessioncode}
\begin{lstlisting}[style=terminal]
cmake -DMODEL=lphi4 ../
make cosmolattice
./lphi4 input=lphi4.in
\end{lstlisting}
\vspace{2mm}

Note that all models distributed publicly in previous versions of \CL{1.X} are still present with their code virtually unchanged.
Similarly, the user can use in \CL{2.0} any model they have previously written themselves for \CL{1.X}, by changing just one line in the file:
\vspace{2mm}

\begin{lstlisting}[style=cppstyle]
MODELNAME(ParameterParser& parser, RunParameters<double>& runPar,
          std::shared_ptr<MemoryToolBox> toolBox):
\end{lstlisting}

\vspace{3mm}
\noindent has to be changed into\\

\begin{lstlisting}[style=cppstyle]
MODELNAME(ParameterParser& parser, RunParameters<double>& runPar,
          auto toolBox):
\end{lstlisting}

\vspace{2mm}
Despite the simplicity of the change above, the backend changes to \CL{2.0} are, in reality, significant. As such, the lattice engine behind \CL{2.0} has been fully rewritten and stands now on its own as a general lattice field theory C++ library: \TempLat \cite{Florio:2026vde}. A more extensive migration guide can be found on the {\tt Migration from v1.X} tab in the \href{http://www.cosmolattice.com}{\color{blue} \cosmolattice website}\,.

The expression algebra as used in \CL{1.X} retains the same user interface, but offers a breadth of new features. Besides a performance improvement of up to 50\% for the same kind of MPI-parallelized CPU setup \CL{1.X} offered (which we report later in Fig.~\ref{fig:scaling-tests}), \CL{2.0} gains the following features from \TempLat:
\begin{itemize}
    \item A new build interface; the user no longer has to manually configure external libraries and link \CL{2.0} to them - the full process has been streamlined and fully automated through CMake. Running just \texttt{cmake .} leads to \CL{2.0} downloading \TempLat and any other dependencies. On demand, \CL{2.0} will also build any external library, including HDF5 and FFTW3 for use with the code.
    \item While \CL{1.X} was only parallelized through MPI, \CL{2.0} offers full hybrid parallelization. This means that while separate nodes are still connected through MPI, locally within a single node as many CPUs or GPUs as available can work together with shared memory, forgoing network overhead fully. The shared-memory parallelization of \TempLat is built on Kokkos \cite{KokkosCore2014,KokkosEcosystem2021, KokkosCore2022}, a prominent performance portability framework.
    \item Performance portability, i.e. the ability to write the same code and then compile and run it for a wide range of hardware, in particular not just CPUs, but also NVIDIA or AMD graphics cards. As current large-scale and exascale clusters increasingly lean towards hybrid infrastructures of CPUs and GPUs, \CL{2.0} is able to take full advantage of large GPU clusters. Lattice simulations are \textit{embarrassingly parallelizable} and \CL{2.0} simulations scale well even at extreme CPU or GPU counts. 
    Therefore, \CL{2.0} numerics are tailor-made for the massive parallelization advantages that GPUs offer. In comparison to pure-CPU computations, GPUs can offer a speed up (on comparable resources) of $10\times$ to $20\times$.
    \item For performance portability and scaling to large clusters, a modern header-only discrete Fourier transformation (DFT) library has been introduced with \TempLat, \ParaFaFT. \ParaFaFT implements a highly efficient distribution algorithm \cite{DBLP:journals/corr/abs-1804-09536} for pencil-decomposition DFTs. \ParaFaFT provides DFTs in arbitrary numbers of dimensions, and supports NVIDIA and AMD hardware, as well as CPUs. In particular, \CL{2.0} drops support for PFFT in favor of \ParaFaFT, which will stay maintained by our developer team, and will experience future improvements.
\end{itemize}

%%%%%%%%%%%%%%%%%%%%%%%%%%%%%%%%%%%%%%%%
\subsection{Scaling tests}
\begin{figure}[t]
   \centering
   \includegraphics[width=\linewidth]{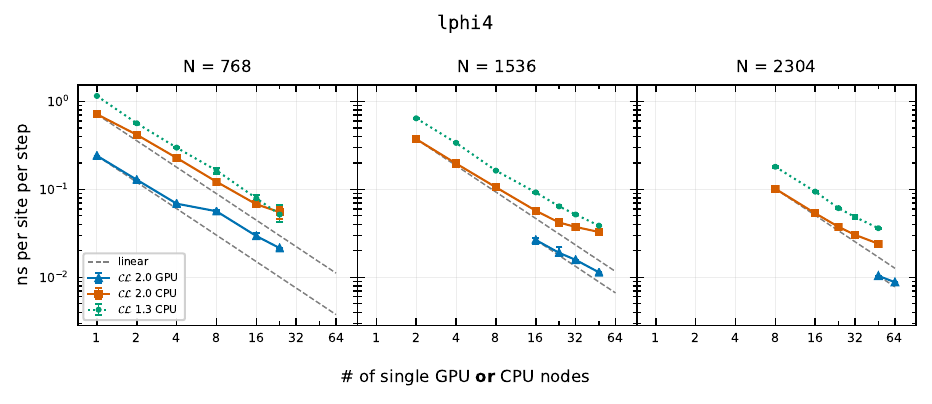}     
   \includegraphics[width=\linewidth]{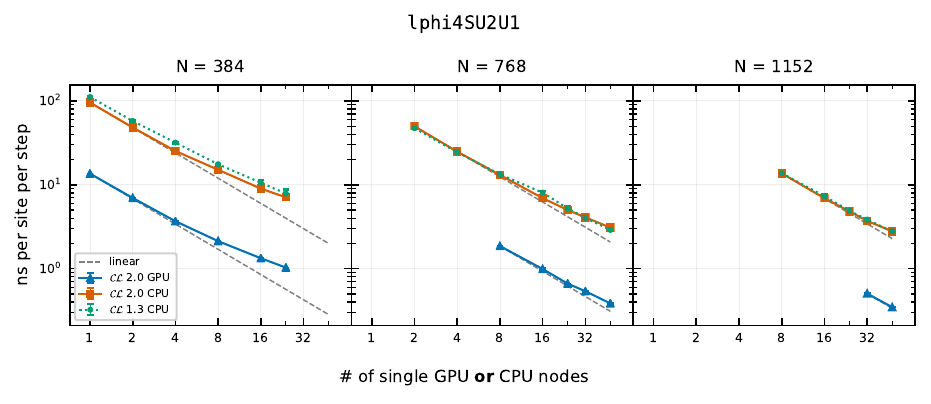}
   \caption{Strong scaling test for model \texttt{lphi4} (above) and \texttt{lphi4SU2U1} (below) with a range of lattice sizes $N$. Each setup has been run for 100 timesteps, with infrequent measurements every 50 steps (power spectra) and frequent measurements every 10 steps (averages). Simulations were performed on the Noctua2 cluster of PC2 \cite{noctua2}, with NVIDIA A100 GPU cards. Grey dashed lines depict perfect scaling.} \label{fig:scaling-tests}
\end{figure}

To show the scalability of both the improved CPU-implementation of \CL{2.0} and the new GPU capabilities, we perform a strong-scaling analysis with \CL{2.0}. To that end, we use two models that are shipped with the code ({\it e.g.}~to simulate scalar field dynamics just after the end of inflation): \texttt{lphi4}, with a {\it mother} field $\phi$ with potential $\propto \phi^4$, and a {\it daughter} field $\chi$ coupled to the former via $\chi^2\phi^2$; and \texttt{lphi4SU2U1}, where the {\it mother} field $\Phi$ is a SU(2) doublet with potential $\propto (\Phi^\dag\Phi)^2$, whereas the {\it daughter} fields are a complex field $\varphi$ coupled via $(\varphi^* \varphi)(\Phi^\dag\Phi)$, and there are also U(1) and SU(2) gauge fields, $A_\mu$ and $C_\mu$, coupled to the scalars via standard gauge interactions as described in Sect.~\ref{subsec:scalargauge}. The \texttt{lphi4} simulation probes the raw throughput efficiency, as the simulation's workload is almost evenly distributed between the halo exchange ({\it i.e.}~{\it ghosting} and thus MPI communication) and actual computation. The \texttt{lphi4SU2U1} simulation, instead, is numerically much heavier and mostly dominated by pure computation, and hence it masks the communication overhead incurred by the MPI parallelization. 

We run every configuration for 100 timesteps and perform infrequent measurements ({\it i.e.}~power spectra) every 50 steps, and frequent measurements ({\it i.e.}~lattice averages for all observables) every 10 timesteps.
All runs have been performed on the Noctua2 HPC cluster~\cite{noctua2} of the Paderborn Center for Parallel Computing. A CPU node is equipped with two AMD EPYC 7763 sockets with 64 Zen3 cores each. GPU nodes have four NVIDIA A100 40 GB GPUs each, connected via NVLink.

{\bf Note -.} Comparisons between GPUs and CPUs are not objective and depend strongly on the hardware configurations chosen for each. To have some measure of relation between the performance of the two, in the following benchmarks, we decided to `equate' a full node of CPUs (with 128 cores in our case) to a single GPU.

In Fig.~\ref{fig:scaling-tests} we show the results of our scaling analysis with lattice sizes $N=768,\,1536$ and $2304$ for \texttt{lphi4} and $N=384,\,768$ and $1152$ for \texttt{lphi4SU2U1}.
In the case of \texttt{lphi4} (upper plots), we see that \CL{2.0} runs $\sim 50\%$ faster than \CL{1.3}. This is simply thanks to the improvements to the backend of \CL{2.0}, \TempLat. When switching  to GPUs, we observe a noticeable speedup by a factor $\sim 5$. For the $N=768$ simulations, the GPU implementation diminishes, however, visibly its efficiency, as we go from 4 to 8 nodes. This is nonetheless expected, as the Noctua2 cluster has 4 GPUs per physical node, as long as a simulation is performed on a single node with multiple GPUs, communication between those is routed through PCIe/NVLink and is extremely fast. Going to multiple physical nodes ({\it i.e.}~8 or more GPUs) incurs additional overhead due to (GPU-aware) MPI communication.

The above scaling can contrasted against the behavior of \texttt{lphi4SU2U1} (lower plots), where the additional MPI overhead is less severe in relation to the time spent in computation. We note here that both the gap between \CL{1.3} and \CL{2.0} reduces on the CPU, but the performance advantage of using GPUs increases, yielding simulations $\sim10$ times faster. As \texttt{lphi4SU2U1} is mainly dominated by the \textit{embarrassingly parallelizable} part of \CL{2.0}, {\it i.e.}~pure computation, we see almost perfect scaling until much larger node counts.

Overall, we find that \CL{2.0} exhibits a good scaling behavior, and significantly improves performance over \CL{1.X} in any kind of computational setup.

%%%%%%%%%%%%%%%%%%%%%%%%%%%%%%%%%%%%%%%%%%%%%%%%%%%%%%%%%%%%%%%%%%%%%%%%%%%%%%%%
\section{Discussion}
\label{sec:Discussion}

The rich landscape of high-energy physics that characterizes  the early Universe, encompasses a broad range of processes that frequently involve non-linear field dynamics, often too intricate to be described reliably using analytical methods alone. Examples include {\it inflation}~\cite{Caravano:2022epk,Figueroa:2023oxc,Figueroa:2024rkr,Sharma:2024nfu,Caravano:2024tlp,Caravano:2024xsb,Jamieson:2025ngu,Barker:2026yyg}; {\it preheating} and other particle-production mechanisms~\cite{Traschen:1990sw, Kofman:1994rk, Shtanov:1994ce, Kaiser:1995fb, Kofman:1997yn, Greene:1997fu, Kaiser:1997mp, Kaiser:1997hg, Greene:1998nh, Greene:2000ew, Peloso:2000hy, Berges:2010zv,Enqvist:2012tc,Figueroa:2015rqa}; the amplification of scalar metric perturbations~\cite{Bassett:1998wg, Bassett:1999mt, Bassett:1999ta, Finelli:2000ya, Chambers:2007se, Bond:2009xx,Linde:2012bt,Imrith:2019njf, Musoke:2019ima, Martin:2020fgl, Adshead:2023mvt,Caravano:2024moy,Caravano:2025diq} that potentially can lead to primordial black hole formation~\cite{Cotner:2019ykd, Martin:2019nuw, GarciaBellido:1996qt, Green:2000he, Cotner:2018vug}; Abelian~\cite{Buividovich:2015jfa,Buividovich:2016ulp,Figueroa:2017hun,Figueroa:2017qmv,Figueroa:2019jsi,Mace:2019cqo,Mace:2020dkp} and non-Abelian~\cite{Akamatsu:2015kau} chiral anomaly dynamics and sphaleron rates~\cite{Philipsen:1995sg,Ambjorn:1995xm,Arnold:1995bh,Arnold:1996dy,Arnold:1997yb,Moore:1997sn,Bodeker:1998hm,Moore:1998zk,Moore:1999fs,Bodeker:1999gx,Arnold:1999uy, Tang:1996qx,Ambjorn:1997jz,Moore:2000mx,DOnofrio:2012phz,DOnofrio:2015gop}; oscillon dynamics~\cite{Gleiser:1993pt,Copeland:1995fq,Amin:2010dc,Amin:2011hj,Gleiser:2011xj,Antusch:2015ziz,Lozanov:2017hjm,Hasegawa:2017iay,Amin:2018xfe,Kitajima:2018zco,Antusch:2019qrr,Ibe:2019lzv,Sang:2019ndv,Kou:2019bbc,Nazari:2020fmk,Sang:2020kpd,Aurrekoetxea:2023jwd,Mahbub:2023faw,Piani:2023aof,Shafi:2024jig,Drees:2025iue,Piani:2025dpy}; phase transitions (first order and others)~\cite{Rajantie:2000fd,Hindmarsh:2001vp,Copeland:2002ku,GarciaBellido:2002aj,Niemi:2018asa,Mazumdar:2018dfl,Hindmarsh:2020hop,Brandenburg:2017neh,Brandenburg:2017rnt}; and cosmic-defect dynamics~\cite{Hindmarsh:1994re, Felder:2000hj, Hindmarsh:2000kd, Rajantie:2001ps, Rajantie:2002dw, Donaire:2004gp, Copeland:2009ga, Hiramatsu:2012sc,Kawasaki:2014sqa,Fleury:2016xrz,Moore:2017ond,Gorghetto:2018myk,Matsunami:2019fss,Saurabh:2020pqe}. Their cosmological consequences may include dark-matter production~\cite{Garcia:2018wtq, Garcia:2021iag, Garcia:2022vwm, Lebedev:2022vwf, Zhang:2023xcd}, primordial magnetogenesis~\cite{DiazGil:2005qp, DiazGil:2007qx, DiazGil:2007dy, DiazGil:2008tf, Fujita:2016qab, Adshead:2016iae, Vilchinskii:2017qul}, baryogenesis~\cite{Kolb:1996jt, Kolb:1998he, GarciaBellido:1999sv, Allahverdi:2000zd, Rajantie:2000nj, Cornwall:2001hq, Copeland:2001qw, Smit:2002yg, GarciaBellido:2003wd, Tranberg:2003gi, Tranberg:2009de, Kamada:2010yz, Lozanov:2014zfa}, modifications to the equation of state after inflation and to cosmic microwave background (CMB) observables~\cite{Podolsky:2005bw,Dufaux:2006ee, Lozanov:2016hid, Figueroa:2016wxr, Krajewski:2018moi, Maity:2018qhi, Antusch:2020iyq, Saha:2020bis, Antusch:2021aiw, Mansfield:2023sqp,Garcia:2023eol,Garcia:2023dyf,Antusch:2025ewc}, and potentially observable gravitational-wave backgrounds~\cite{Khlebnikov:1997di,Easther:2006gt,Easther:2006vd,Garcia-Bellido:2007nns,GarciaBellido:2007af,Dufaux:2007pt,Dufaux:2008dn,Dufaux:2010cf,Figueroa:2012kw,Hiramatsu:2013qaa,Hindmarsh:2013xza,Zhou:2013tsa, Bethke:2013aba,Bethke:2013vca,Figueroa:2014aya,Hindmarsh:2015qta,Figueroa:2016ojl,Antusch:2016con, Hindmarsh:2017gnf,Antusch:2017flz,Antusch:2017vga,Figueroa:2017vfa,Cutting:2018tjt, Liu:2018rrt,Lozanov:2019ylm, Adshead:2019lbr,Adshead:2019igv, Cutting:2019zws,Pol:2019yex, Figueroa:2020lvo,Cutting:2020nla,Figueroa:2022iho,Cosme:2022htl, Klose:2022knn,Cui:2023fbg,Baeza-Ballesteros:2023say,Servant:2023tua,Baeza-Ballesteros:2024sny,Caravano:2026hca}. 

Developing powerful, flexible and robust numerical techniques to simulate nonlinear field dynamics is therefore essential to obtain a solid understanding of early Universe phenomena, particularly in the present era of precision observational cosmology. {\bf Lattice Cosmology Techniques} (LCT) have emerged as a powerful framework for understanding non-linear field dynamics, and are expected to play an increasingly important role in shaping observational strategies for probing the early Universe. In such a context, \cosmolattice (\CLns) represents a modern computational tool that enables the exploration of complex non-linear regimes that were beyond practical reach only a short time ago. In this paper we have introduced \CL{2.0}, which substantially extends the physics scope and computational capabilities of previous versions of the code. An extensive documentation on the use of the code is provided on the \href{http://www.cosmolattice.com}{\color{blue} \cosmolattice website}, from where the code can be downloaded. Alternatively, visit our \href{https://github.com/cosmolattice/cosmolattice}{\color{blue} GitHub repository}\,.

\CL{2.0} significantly broadens the physics scope of the code by incorporating lattice implementations of non-canonical interactions. These include scalar fields non-minimally coupled to gravity through interactions of the form $\phi^2R$, as well as axion-like fields coupled to Abelian gauge sectors through $\phi F_{\mu\nu}\widetilde F^{\mu\nu}$. The new version also supports the preparation and evolution of specialized field configurations, including scaling networks of cosmic strings and domain walls, together with techniques for preserving defect-core resolution. Furthermore, scalar-field dynamics can be simulated on reduced $(1+1)$- and $(2+1)$-dimensional lattices, enabling computationally cheaper approximations to statistically isotropic three-dimensional systems.

On the technical side, \CL{2.0} introduces standard and low-storage Runge--Kutta integrators, providing stable evolution schemes for non-canonical, non-Hamiltonian, dissipative, or potentially stiff systems that are not well suited to symplectic methods. It also incorporates optimized GW evolution, flexible initialization from arbitrary power spectra, and extended output capabilities for field amplitudes, energy densities, and user-defined lattice subsets. Finally, GPU support provides substantial performance gains, accelerating representative simulations by factors of $\mathcal{O}(10)$ relative to CPU execution while retaining the parallel and modular structure of previous versions of the code.

In upcoming \CL{2.X} upgrades, expected in the near future, the code will be further extended to include non-minimal kinetic scalar theories of the form $\mathcal{G}_{ab}(\lbrace\phi_c\rbrace)\partial_\mu\phi^a\partial^\mu\phi^b$, ALP-SU(2) interactions of the type $\phi G_{\mu\nu} \tilde G^{\mu\nu}$, and possibly other features. Furthermore, complementing our current LCT reviews \ArtI~\cite{Figueroa:2020rrl} and \ArtII~\cite{Baeza-Ballesteros:2025tme}, a third entry in the monographic series, \ArtIII~\cite{Figueroa:2026esg} has been also posted on ArXiv. There, we discuss the theoretical basis of new modules planned for \CL{3.0}, where we introduce lattice formulations of relativistic and non-relativistic fluids, in isolation or interacting with scalar and/or gauge fields. In the medium term, we also plan to complete the LCT monograph series with \ArtIV, expected to provide the theoretical foundation for \CL{4.0}, where we will consider lattice formulations of general relativity sourced by scalar, gauge, and fluid {\it dof}.

\begin{center}
    ---------------
\end{center}

{\bf Acknowledgements -.} The work of DGF (0000-0002-4005-8915) is supported by the grants PROMETEO/2021/083, EUR2022-134028, PID2023-148162NB-C22, PRTR-C17.I01, and ASFAE/2022/020. FT (0000-0003-1883-8365) acknowledges support through the \textit{Atracción de Talento César Nombela} fellowship No 2025-T1/COM-36104 funded by Comunidad de Madrid (Spain). NL thanks Josef Dvořáček for his technical support in preparing the simulations for GPUs, and acknowledges support by the Czech Science Foundation, GAČR, Project No. 24-13079S.  AU
(0000-0002-0238-8390) acknowledge support from Eusko Jaurlaritza IT1628-22, from PID2024-156016NB-I00 grant funded by MCIN/AEI/10.13039/501100011033/ and by ERDF: “A way of making Europe”, and the University of the Basque
Country grant PIF20/151, and thanks in particular the Institute for Theoretical Physics at the University of Münster for hosting him during the final months of this project. AF and FS  are funded
by the Deutsche Forschungsgemeinschaft (DFG, German Research Foundation) through the Emmy
Noether Programme Project No. 545261797.

Computations for this work were performed on the {\tt Graviton} cluster of the SOM group at the Instituto de Física Corpuscular (IFIC), the ARINA and Solaris clusters at the University of the Basque Country (UPV/EHU), the Hyperion cluster
from the DIPC Supercomputing Center, the MareNostrum 5 cluster at Barcelona Supercomputing Center (BSC), the FinisTerrae III cluster at Centro de Supercomputación de Galicia (CESGA), the Lluis Vives and Tirant II clusters at the University of Valencia, the Phoebe Cluster at CEICO/FZU. We also acknowledge support of the IT resources of the UC3M C3 Cluster, co-financed through action EQC2021-007184-P. We acknowledge as well the EuroHPC Joint Undertaking for awarding this project access to the EuroHPC supercomputer LUMI, hosted by CSC (Finland) and the LUMI consortium through a EuroHPC Regular Access call.
We also acknowledge the computing time made available to us on the high-performance computer Noctua2 at the NHR Center Paderborn Center for Parallel Computing (PC2). This center is jointly supported by the Federal Ministry of Research, Technology and Space and the state governments participating in the National High-Performance Computing (NHR) joint funding program (\url{www.nhr-verein.de/en/our-partners}).

%%%%%%%%%%%%%%%%%%%%%%%%%%%%%%%%%%%%%%%%%%%%%%%%%%%%%%%%%%%%%%%%%%%%%%%%%%%%%%%%

\footnotesize
\bibliography{biblio,extra}
\bibliographystyle{elsarticle-num}

\end{document}